\documentclass[twocolumn,twocolappendix]{aastex631}
\usepackage{graphicx}
\usepackage[version=4]{mhchem}
\usepackage{subfigure}
\usepackage[utf8]{inputenc}
\usepackage{hyperref}
\usepackage{amsmath}
\usepackage{array}
\usepackage{xspace}

\newcommand\numberthis{\addtocounter{equation}{1}\tag{\theequation}}
\newcommand\POSEIDON{\texttt{POSEIDON}\xspace}

\newcommand{\correction}[1]{\textcolor{black}{#1}}

\defcitealias{Lustig-Yaeger2023}{Lustig-Yaeger \& Fu et al.}
\defcitealias{Moran2023}{Moran \& Stevenson et al.}
\defcitealias{May2023}{May \& MacDonald et al.}

\received{February 29, 2024}
\revised{April 3, 2025}
\accepted{April 7, 2025}
\shorttitle{High-Resolution Retrievals with \POSEIDON}

\begin{document}

\title{Open Source High-Resolution Exoplanet Atmosphere Retrievals with \POSEIDON}

\correspondingauthor{Ruizhe Wang}
\email{rwang3@caltech.edu}

\author[0009-0007-8482-2704]{Ruizhe Wang}    
\affiliation{Division of Geological	and	Planetary Sciences, California Institute of Technology, Pasadena, CA 91125,	USA}
\affiliation{Department of Astronomy and Carl Sagan Institute, Cornell University, 122 Sciences Drive, Ithaca, NY 14853, USA}

\author[0000-0003-4816-3469]{Ryan J. MacDonald}
\altaffiliation{NHFP Sagan Fellow}
\affiliation{Department of Astronomy, University of Michigan, 1085 S. University Ave., Ann Arbor, MI 48109, USA}
\affiliation{Department of Astronomy and Carl Sagan Institute, Cornell University, 122 Sciences Drive, Ithaca, NY 14853, USA}

\author[0000-0002-9308-2353]{Neale P. Gibson}
\affiliation{School of Physics, Trinity College Dublin, University of Dublin, Dublin-2, Ireland}

\author[0000-0002-8507-1304]{Nikole K. Lewis}
\affiliation{Department of Astronomy and Carl Sagan Institute, Cornell University, 122 Sciences Drive, Ithaca, NY 14853, USA}

\begin{abstract}

High-resolution spectroscopy (R $>$ 25,000) has opened new opportunities to characterize exoplanet atmospheres from the ground. By resolving individual lines in planetary emission and transmission spectra, one can sensitively probe the chemical inventory and temperature structure of exoplanets. However, a significant challenge to reliable and reproducible atmospheric inferences from high-resolution datasets has been the lack of open source codes for high-resolution retrievals. Here, we present a unified high-resolution retrieval framework, for both emission and transmission spectroscopy, made publicly available within the open source \POSEIDON retrieval code. Our high-resolution retrieval framework is fast (typically $<$ 12 hours), accessible (no GPUs required), and well-documented via Python notebooks. We validate our framework by reproducing previous emission retrievals of the hot Jupiter WASP-77Ab and transmission retrievals of the ultra-hot Jupiter WASP-121b. Our results are broadly consistent with those of published works when making the same data detrending assumptions, but we demonstrate that user choices can subtly propagate into retrieved chemical abundances. 



\end{abstract}

\section{Introduction}
\label{sec: introduction}

Transmission and emission spectroscopy are fundamental methods for characterizing exoplanet atmospheres, enabling remote observers to infer their chemical composition and atmospheric properties. Many early breakthroughs in exoplanet spectroscopy were realized with space-based low-resolution spectroscopy (LRS, $R=\lambda/\Delta \lambda \leq 200$), primarily with the Hubble Space Telescope (e.g., \citealt{Charbonneau2002, Deming2013, Knutson2014}) and the Spitzer Space Telescope (e.g., \citealt{Grillmair2008, Kreidberg2014b}). Recently, the JWST has opened new frontiers in ultra-precise LRS of exoplanet atmospheres, ranging from giant planets (e.g., \citealt{JWSTTransitingExoplanetCommunityEarlyReleaseScienceTeam2023,Coulombe2023,Grant2023,Bell2024,Welbanks2024}) to terrestrial worlds (e.g., \citetalias{Lustig-Yaeger2023} \citeyear{Lustig-Yaeger2023}; \citealt{Lim2023}; \citealt{Hu2024}). 

Ground-based high-resolution spectroscopy (HRS, $R \geq 20,000$) has emerged in parallel to space-based LRS as a formidable technique for exoplanet characterization (e.g., \citealt{Snellen2010,Birkby2013,Hoeijmakers2018b,Brogi2019,Ehrenreich2020,Gandhi2022,Pelletier2023,Smith2024}). At high spectral resolution, molecular, atomic, or ionic transitions are resolved as lines, revealing the Doppler shift of planetary emission and transmission signatures. HRS can, therefore, shed light on different physical regimes and atmospheric processes compared to LRS. Since most individual absorption lines are not detectable with present instruments, high-resolution studies of exoplanet atmospheres have traditionally relied on cross-correlating template models with HRS observations to boost the signal-to-noise ratio. Observations at high resolution can currently only be accomplished with large ground-based telescopes, which consequently must contend with telluric contamination and a loss of continuum information from data filtering. However, despite these issues, ground-based HRS has delivered stunning insights into the chemical inventories (e.g., \citealt{Hoeijmakers2019, Merritt2021,Kesseli2022,Pelletier2023}), pressure-temperature profiles (e.g., \citealt{Brogi2019,Wardenier2023}), and atmospheric dynamics (e.g., \citealt{Snellen2010, Brogi2016}) of dozens of hot gas giant exoplanets. Table \ref{tab:instruments} lists some of the major observatories and instruments that have detected exoplanet atmospheres using HRS.

Measuring the chemical composition of exoplanet atmospheres via HRS has been a key focus of the field. While the classical cross-correlation approach has proven effective in identifying chemical species, we ultimately seek constraints on the range of chemical abundances, temperature-pressure profiles, and other atmospheric properties. Deriving statistical constraints on the atmospheric properties consistent with a given dataset --- called \emph{atmospheric retrieval} \citep[e.g.][]{Madhusudhan2018} --- is challenging for HRS due to the loss of continuum information, the resultant distortion of the underlying planet signal from filtering, and the difficulty in defining a unique way to compare model spectra to data. \citet{Brogi2019} introduced the first Bayesian atmospheric retrieval method for High-Resolution Cross-Correlation Spectroscopy (HRCCS), demonstrating the technique on hot Jupiter emission spectra. By relating cross-correlation functions to statistical likelihood functions, HRCCS retrieval techniques yield probability distributions on the model parameters. \citet{Gibson2020} developed a similar technique for HRCCS of exoplanet transmission spectra. In the years since, HRCCS retrievals have been applied to many hot Jupiters, including \correction{emission retrievals of HD~209458b \citep{Brogi2019,Gandhi2019}, HD~189733b \citep{Brogi2019,Finnerty2024}, WASP-189b \citep{Yan2020,Lesjak2024}, WASP-77Ab \citep{Line2021, Smith2024}, $\tau$~boo~b \citep{Pelletier2021,Panwar2024}, WASP-18b \citep{Brogi2023}, MASCARA-1b \citep{Ramkumar2023}, and WASP-121b \citep{Pelletier2025}, alongside transmission retrievals of WASP-121b \citep{Gibson2020,Gibson2022,Gandhi2023,Maguire2023}, HD~189733b \citep{Boucher2021,Klein2024,Blain2024a}, and WASP-76b \citep{Pelletier2023,Maguire2024,Gandhi2024}.}

HRCCS retrieval techniques have seen limited application compared to space-based LRS datasets, despite the abundance of HRS exoplanet data and the rich information contained within. This is primarily due to two reasons: (i) unlike LRS applications (for which over 50 retrieval codes exist; \citealt{MacDonald2023b}), very few retrieval codes can handle the unique complexities of HRCCS retrievals; and (ii) there is a severe lack of publicly available, well-documented, codes for HRCCS retrievals. \citet{Line2021} provide a link to a Dropbox with a Graphics Processing Unit (GPU) retrieval code for the emission spectrum of WASP-77Ab, while \citet{Klein2024} and \cite{Debras2024} present an open source HRCCS retrieval code for transmission spectroscopy with the SPIRou instrument. \correction{Very recently, \citet{Blain2024b} introduced a framework for high spectral resolution retrievals within petitRADTRANS \citep{Molliere2019}.} The LRS community has benefited from intercomparisons between open source codes, which help to understand the impact of different modeling techniques and assumptions. For example, \citet{Barstow2022} compared five different codes and found that small model differences can lead to noticeable differences in the retrieval results. Comparative retrieval studies for HRCCS are essential to understand the retrieved results. The lack of availability of open source HRCCS retrieval codes is a significant barrier to progress in this rapidly moving area of exoplanet science. 

Here, we introduce our open source implementation of HRCCS retrievals for both emission and transmission spectra within the \POSEIDON retrieval code \citep{MacDonald2017,MacDonald2023}. Our retrieval framework builds upon the methods described in \citet{Brogi2019} for emission spectroscopy and in \citet{Gibson2020} and \citet{Gibson2022} for transmission spectroscopy. We validate our framework via multiple injection and recovery tests before running HRCSS retrievals on HRS observations of WASP-77Ab and WASP-121b. We further experiment with different stages of the pipeline (e.g., data filtering, model filtering, and likelihood mapping) to demonstrate the impact on the final retrieval results. Documentation, tutorials, and example notebooks can be found on \POSEIDON's website\footnote{\href{https://poseidon-retrievals.readthedocs.io/en/latest/index.html}{\POSEIDON's documentation.}}. \correction{Our implementation of high-resolution retrievals is publicly available as of \POSEIDON v1.3.}

Our paper can also serve as a standalone review of high-resolution retrieval methods for both emission and transmission spectra. In Section \ref{sec: methods}, we describe the HRCCS retrieval methodology for emission and transmission spectra. In Section \ref{sec: observations}, we describe the observations used in this paper, including the data reduction and preprocessing steps. Subsequently, in Section \ref{sec: injection}, we demonstrate our retrieval framework on simulated data with injected signals to validate \POSEIDON's HRCCS capabilities for realistic datasets. Our results are presented in Section \ref{sec: WASP77} and Section \ref{sec: WASP121}, where we apply \POSEIDON to HRS observations of WASP-77Ab in emission and WASP-121b in transmission. Finally, we summarize our results and discuss the implications in Section \ref{sec: summary}.

\begin{deluxetable*}{ccccc}[ht!]
\label{tab:instruments}
\tablecaption{A non-exhaustive list of high-resolution instruments that have been used for high-resolution cross-correlation analysis}
\tablehead{
    \colhead{Observatory} & \colhead{Instrument} & \colhead{WL (nm)} & \colhead{R ($10^3$)} & \colhead{Example References}
}
\startdata
VLT & CRIRES / & 1000-5000 & $\sim$ 100 & \cite{Kaeufl2004}*, \cite{Snellen2010}, \cite{deKok2013}, \\ 
& CRIRES$^{+}$ & & & \cite{Brogi2017}, \cite{Cabot2019}, \cite{Ramkumar2023} \\
VLT & \textbf{UVES (blue)} & 300-500 & $\sim$ 80 & \cite{D'Odorico2000}*, \cite{Gibson2020} \\
VLT & \textbf{UVES (red)} & 420-1100 & $\sim$ 110 & \cite{Merritt2020}, \cite{Gibson2022} \\
VLT & ESPRESSO & 380-788 & $\sim$ 70/140/190 & \cite{Pepe2021}*, \cite{AzevedoSilva2022}, \cite{Maguire2023} \\
Gemini North & MAROON-X & 500-920 & $\sim$ 80 & \cite{Seifahrt2022}*, \cite{Pelletier2023} \\
Gemini South & \textbf{IGRINS} & 1450-2450 & $\sim$ 45 & \cite{Park2014}*, \cite{Flagg2019}, \cite{Line2021} \\
Gemini South & GRACES & 400-1000 & $\sim$ 60 & \cite{Chene2014}*, \cite{Deibert2021}, \cite{Flagg2023} \\
\correction{Gemini South} & \correction{GHOST} & \correction{383-1000} & \correction{$\sim$ 76} & \cite{Kalari2024}*, \cite{Deibert2024}, \cite{Langeveld2025} \\
Keck & NIRSPEC & 950-5400 & $\sim$ 25 & \cite{McLean1998}*, \cite{Piskorz2016}, \cite{Buzard2020} \\
La Silla & HARPS & 383-690 & $\sim$ 115 & \cite{Mayor2003}*, \cite{Stangret2022}, \cite{Bourrier2020} \\
TNG & HARPS-N & 378-691 & $\sim$ 115 & \cite{Cosentino2012}*, \cite{Langeveld2022} \\
TNG & GIANO & 950-2500 & $\sim$ 50 & \cite{Origlia2014}*, \cite{Brogi2018}, \cite{Giacobbe2021} \\
LBT & PEPSI & 383-907 & $\sim$ 250 & \cite{Strassmeier2015}*, \cite{Johnson2023}, \cite{Petz2023} \\
Calar Alto & CARMENES & 520-1710 & $\sim$ 80/94 & \cite{Quirrenbach2014}*, \cite{Alonso-Floriano2019} \\
Subaru & HRD & 300-1000 & $\sim$ 160 & \cite{Noguchi2002}*, \cite{Nugroho2017} \\
Subaru & IRD & 970-1750 & $\sim$ 70 & \cite{Kotani2018}*, \cite{Nugroho2021} \\
CFHT & SPIRou & 980-2350 & $\sim$ 75 & \cite{Artigau2014}*, \cite{Ridden-Harper2023} \\
WIYN & NEID & 380-930 & $\sim$ 100 & \cite{Schwab2016}*, \cite{Yang2024} \\
\enddata
\tablecomments{Data from \textbf{bolded} instruments are used in this study. Instrument references are denoted by `*'. `WL' is wavelength. $R$ is the spectral resolution ($\lambda/\Delta \lambda$).}
\end{deluxetable*}

\section{High-Resolution Retrieval Methodology in \POSEIDON}\label{sec: methods}

\correction{Here, we introduce the addition of high-resolution atmospheric retrieval techniques for ground-based high-dispersion spectrographs to the open source \POSEIDON retrieval framework. We first describe modifications to the forward model for high-resolution retrieval applications, before describing our approach to data detrending and high-resolution retrieval.}

\subsection{\texorpdfstring{High-Resolution Model Planetary Spectra}{Modeling the Planetary Atmosphere (The Forward Model)}}
\label{sec: forward model}

We use the \POSEIDON package \citep{MacDonald2017,MacDonald2023} to model and retrieve exoplanet transmission and emission spectra. \POSEIDON's transmission forward model is extensively described in \citet{MacDonald2022} while the emission forward model is described in \citet{Coulombe2023} and \citet{Mullens2024}. \POSEIDON has been widely used to analyze low-resolution exoplanet spectra from \emph{Hubble} \citep[e.g.,][]{Kilpatrick2018,Rathcke2021,Alderson2022,Mullens2024,Boehm2025} and JWST (e.g., \citealt{Taylor2023}, \citealt{Fournier-Tondreau2023}; \citetalias{May2023} \citeyear{May2023}; \citealt{Gressier2024}; \citealt{Piaulet-Ghorayeb2024}). \POSEIDON is publicly available as a Python package on GitHub\footnote{\href{https://github.com/MartianColonist/POSEIDON}{\POSEIDON's GitHub repository.}}.

To adapt \POSEIDON for HRCCS retrievals, we must first decide on the space of retrieval model assumptions. Ultimately, there is a trade-off between model complexity and computational runtime. First, we note that for high-resolution applications, many authors have highlighted the sensitivity to multidimensional atmospheric properties. For example, \citet{Beltz2021} demonstrates that the 3D atmospheric structure of hot Jupiters has a first-order influence at high resolution, while \citet{Wardenier2023} concludes that observing the phase-resolved absorption signal of multiple species is key to constraining the 3D thermochemical structure and dynamics of ultra-hot Jupiters. \citet{Gandhi2022} retrieves asymmetries between morning and evening terminators with day-night winds, using two separate T-P profiles with independent atmospheric parameters and a broadening kernel to extract spatial information. They find their 2D model to be statistically favored over traditional 1D models, and thus showcase the potential of revealing multidimensional features with current and future facilities. While \POSEIDON does support multidimensional transmission spectra retrievals, which allows day-night variations of temperature and chemical abundances (see \citealt{MacDonald2022}), for this initial study we restrict our focus to 1D atmospheric structures to enable a fair comparison to previous HRCCS retrieval studies. Future studies can investigate the degree to which signatures of 3D effects can be retrieved --- especially relative to other significant uncertainties, such as data filtering. Second, we assume vertically constant gas volume mixing ratios, neglecting the potential for strong gradients over the pressure range probed by the spectra. These choices keep the parameter space defining the forward model computationally tractable, given the need to calculate $\gtrsim 10^5$ model spectra at R$\gtrsim$100,000 during a typical retrieval.

We restrict our focus here to hydrogen-dominated gaseous planets. Each model atmosphere contains several trace gases, molecules and atoms, occupying a fraction of the atmosphere given by the mixing ratio, $X_{\mathrm{i}} \equiv n_{\mathrm{i}} / n_{\mathrm{tot}}$, where $n_{\mathrm{i}}$ is the number density of the i$^{\mathrm{th}}$ gas. The remaining bulk composition is H$_2$ and He with a fixed ratio of $n_{He}$ / $n_{H_2}$ = 0.17. For emission spectra, we model the atmosphere from 10$^{-5}$--100\,bar with 100 layers spaced linearly in log-pressure. For transmission spectra, we extend the upper boundary of the atmosphere to 10$^{-12}$\,bar to avoid line saturation/clipping for strong line cores. We set the reference pressure to be 0.01\,bar, unless specified otherwise, where this corresponds to the pressure of the literature planetary `white light' radius, $R_p$. We employ several parameterizations for the atmospheric pressure-temperature (P-T) profile. While isothermal P-T profiles typically suffice for transmission spectra, we use the more flexible P-T profile from \citet{Madhusudhan2009} for emission spectra (given the greater sensitivity to temperature gradients in emission spectra). We can optionally include parameterized aerosols, including an optically thick cloud deck and scattering haze.

\begin{figure*}[ht]
\centering
\begin{subfigure}
  \centering
  \includegraphics[width=0.49\linewidth]{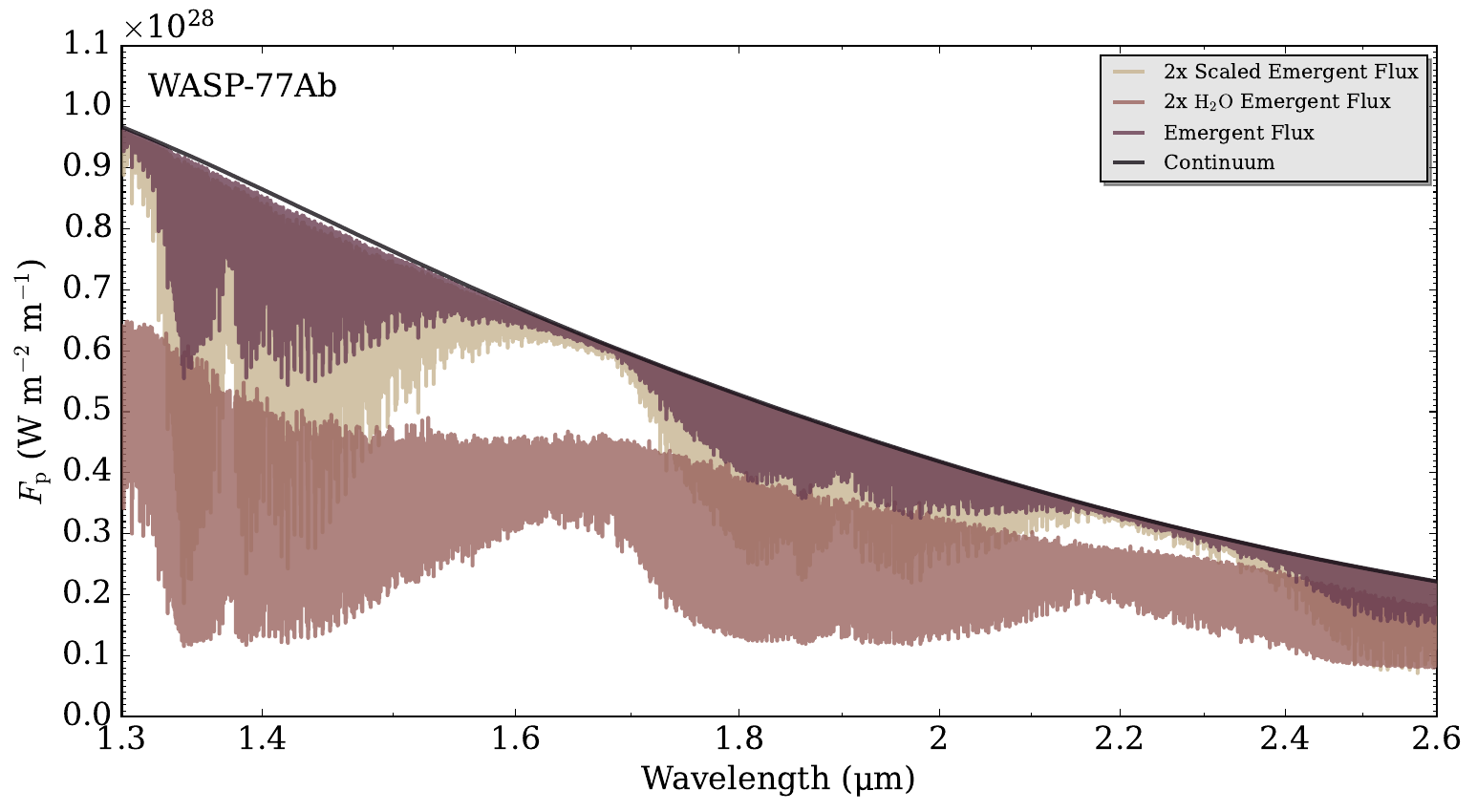}
\end{subfigure}
\begin{subfigure}
  \centering
  \includegraphics[width=0.49\linewidth]{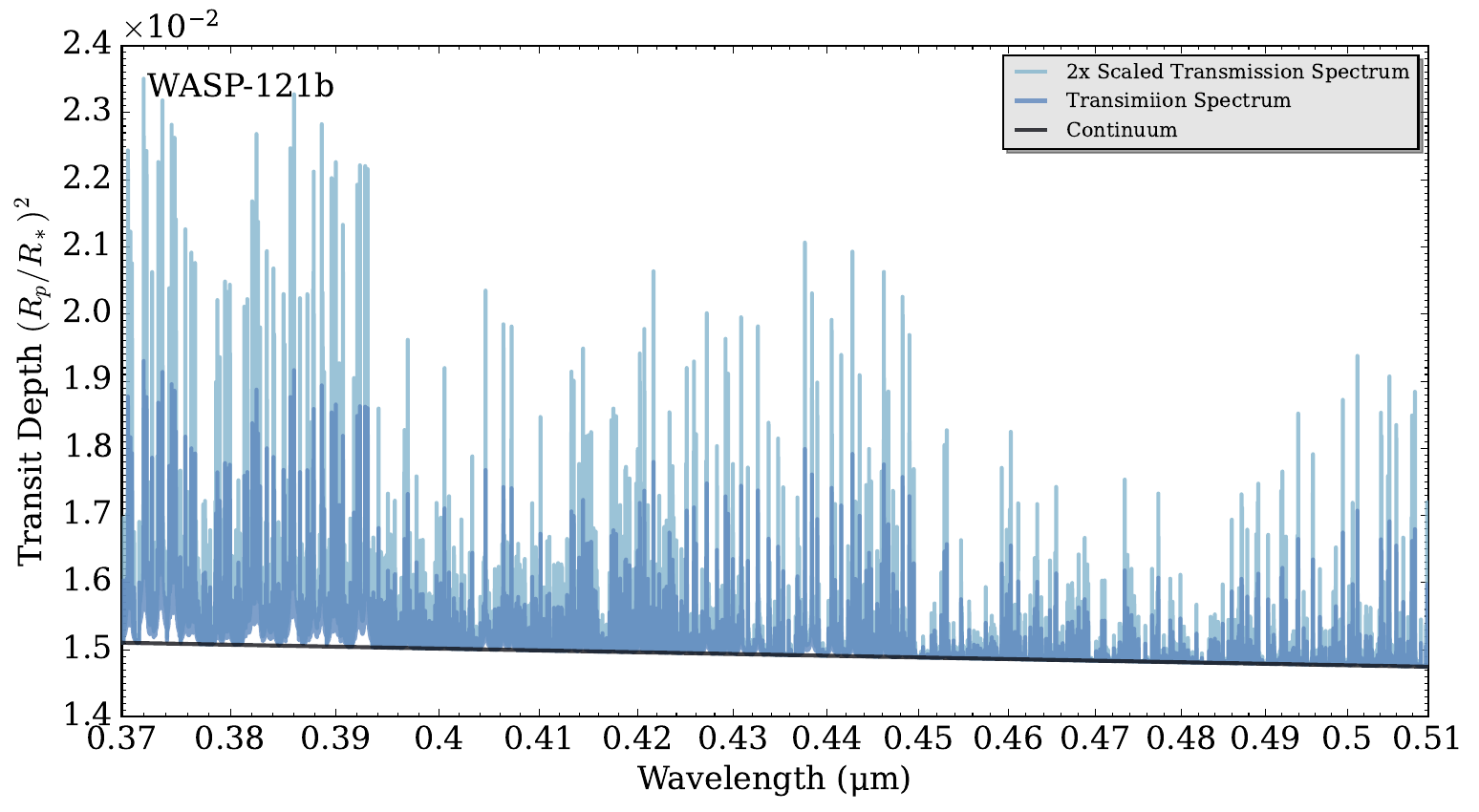}
\end{subfigure}
\begin{subfigure}
  \centering
  \includegraphics[trim=0cm 0.2cm 0cm 0cm, clip, width=0.49\linewidth]{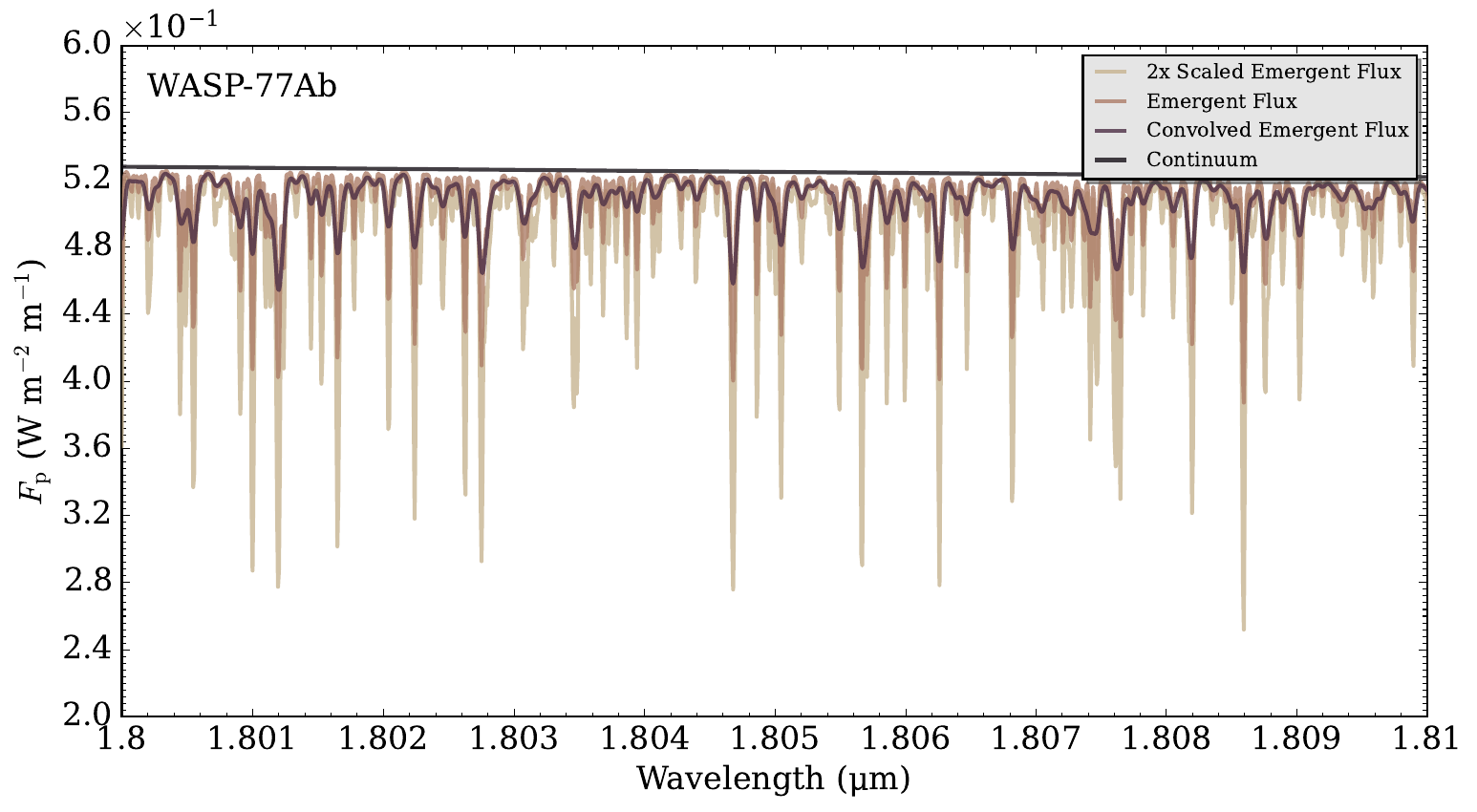}
\end{subfigure}
\begin{subfigure}
  \centering
  \includegraphics[trim=0cm 0.2cm 0cm 0cm, width=0.49\linewidth]{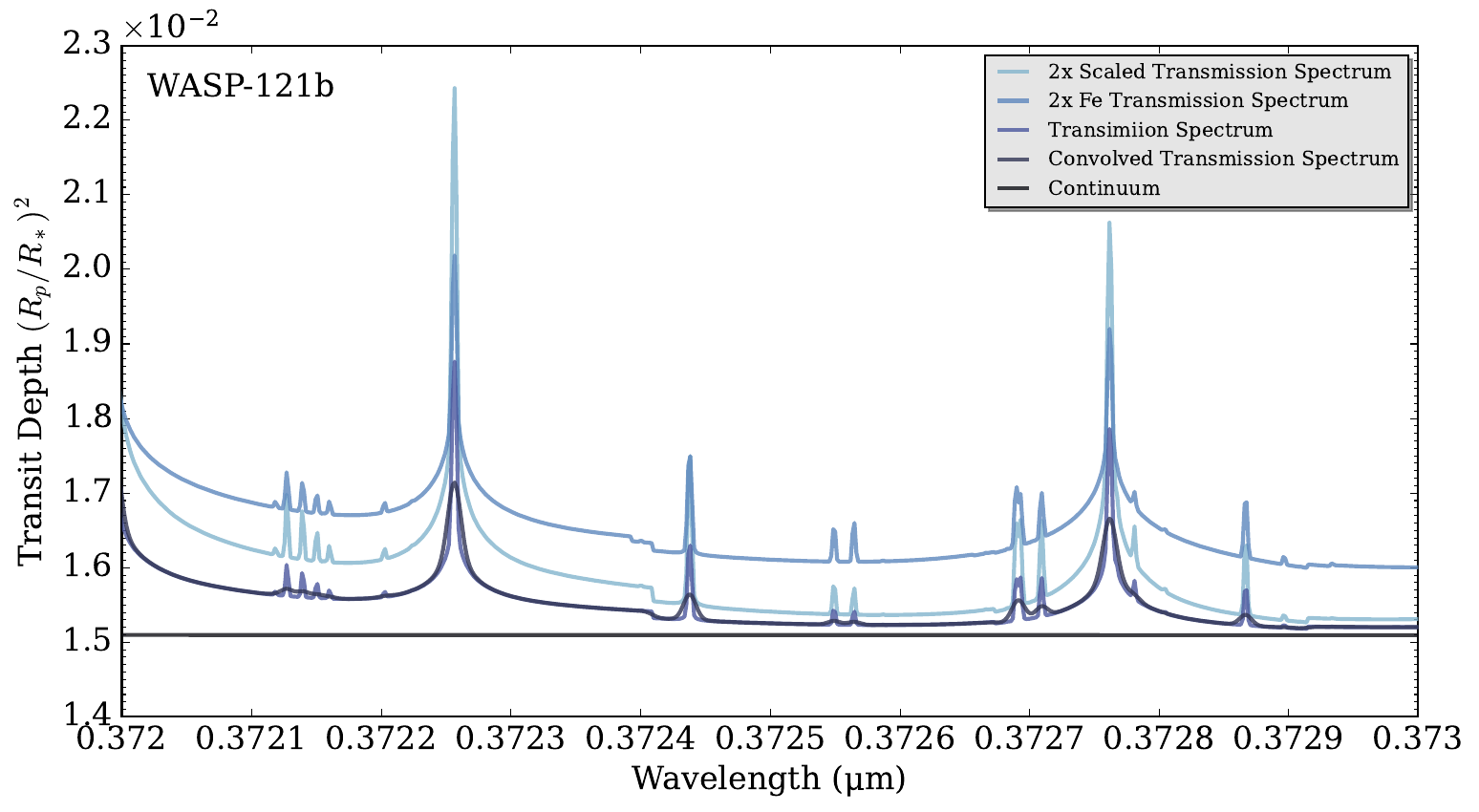}
\end{subfigure}
\caption{Example high-resolution transmission and emission forward models. Left: emergent flux with $\log X_{\ce{H2O}}=-6$. Right: transmission spectrum with $\log X_{\ce{Fe}}=-6$. Models scaled by $\alpha=2$ and models with doubled abundances are over-plotted for comparison. Models convolved by a Gaussian kernel with width 3 are also plotted.}
\label{fig: forward_model}
\end{figure*}

Figure~\ref{fig: forward_model} shows example high-resolution ($R = 250,000$) emission and transmission spectra computed by \POSEIDON. Emergent emission spectra, $F_p$, are calculated in units of W\,m$^{-2}$\,m$^{-1}$, while transmission spectrum are dimensionless with units of $(R_{\mathrm{p, \, effective}}/R_*)^2$. To account for the scale factor $\alpha$ in injection tests (defined below), we compute the continuum without any chemical opacity, subtract the continuum from the forward model, multiply by $\alpha$, and then finally add back the continuum.

\begin{figure*}[t!]
\centering
\begin{subfigure}
  \centering
  \includegraphics[trim=5 5 5 5, clip, width=0.47\linewidth]{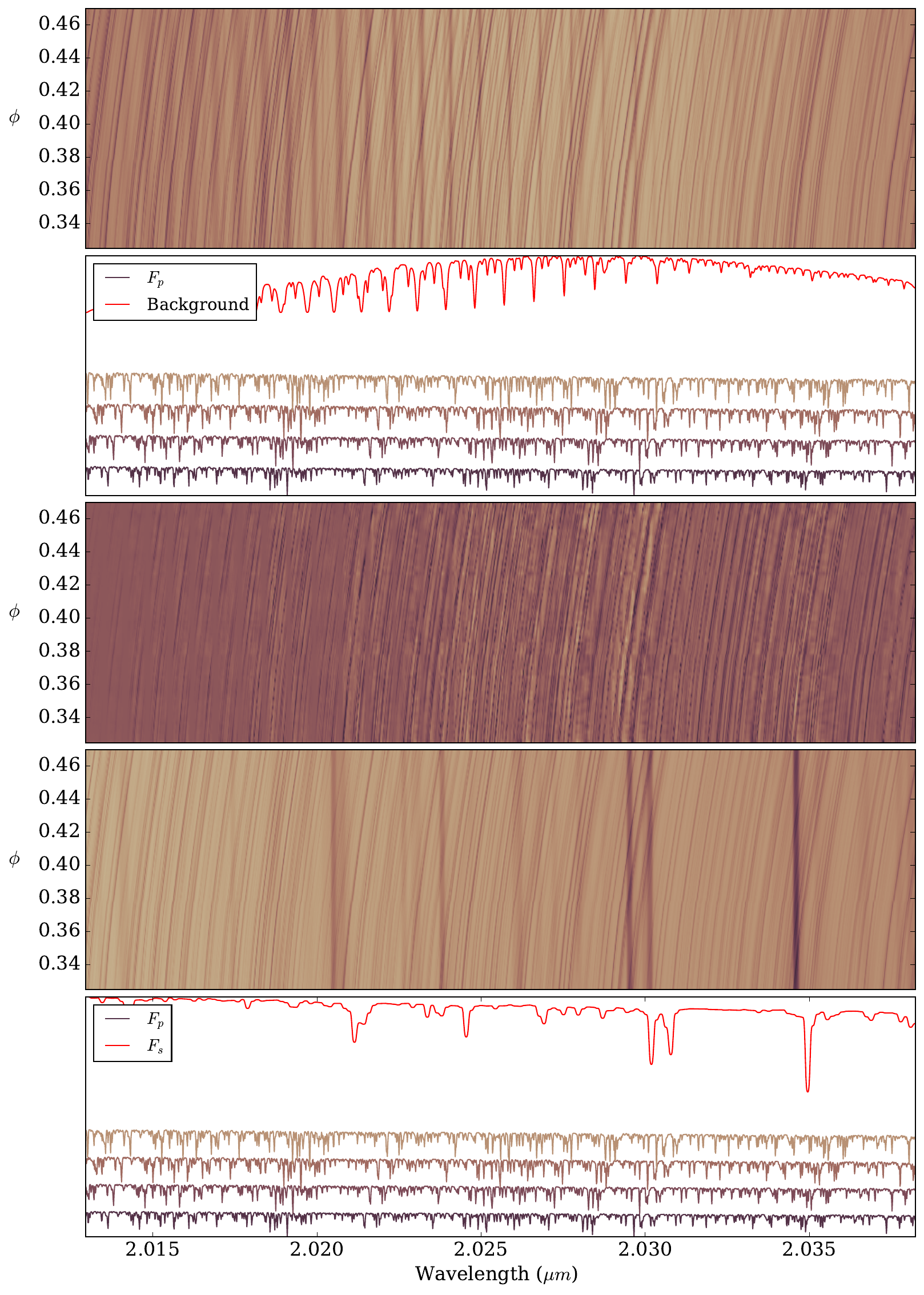}
  \hspace{5mm}
\end{subfigure}
\begin{subfigure}
  \centering
  \includegraphics[trim=5 5 5 5, clip, width=0.471\linewidth]{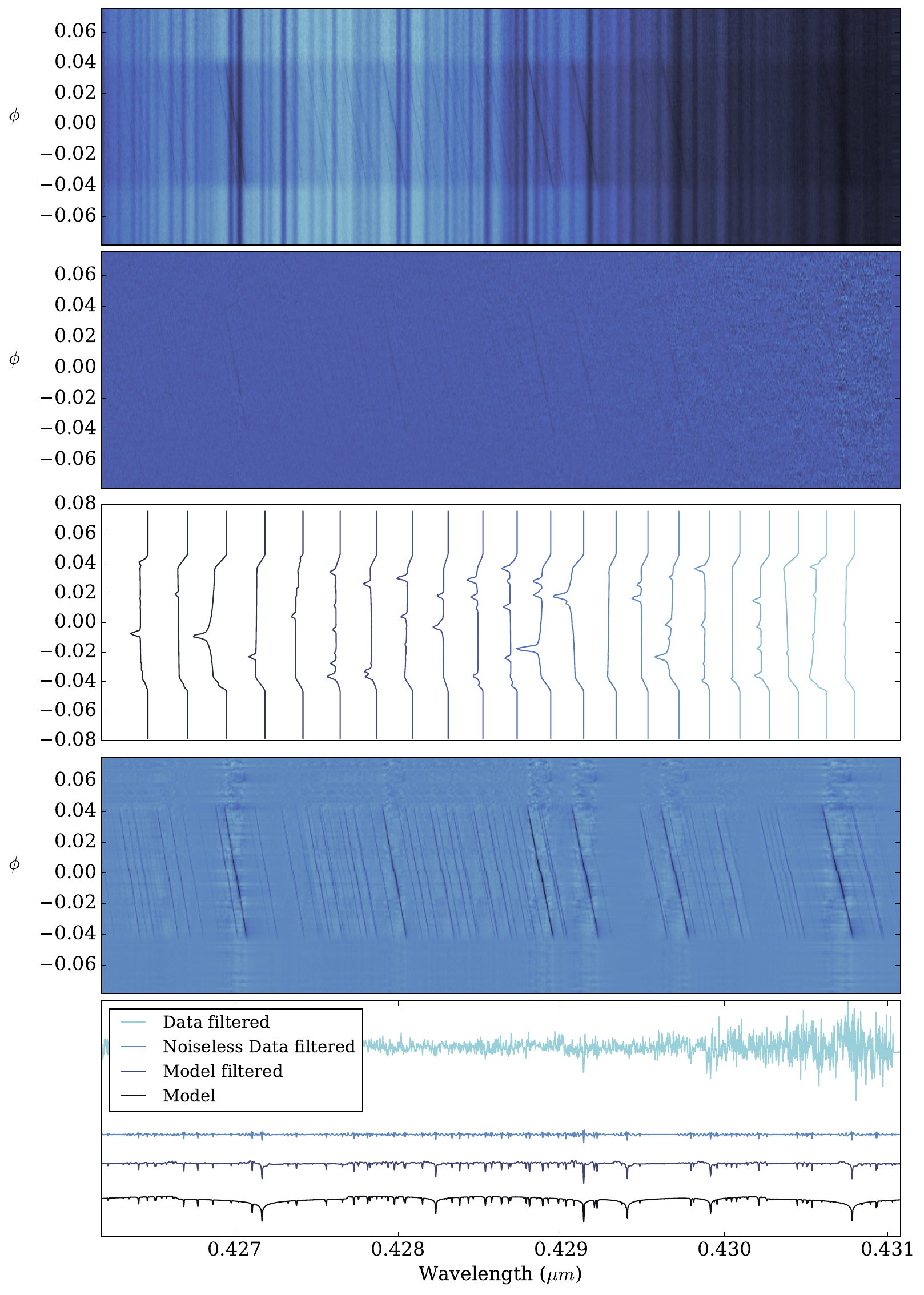}
\end{subfigure}
\caption{\correction{Panel 1: The observation injected with an enhanced signal (to be seen by eye for demonstration purposes). In this paper, red color theme is used for emission spectroscopy (left column) and blue is used for transmission spectroscopy (right column).} Panel 2 left: Red consists of the background contribution from star, noise, and instrumental effects. $F_p$ are injected planet signals at different phases. They highlight the wavelength-shifting nature of planet signals. Panel 2 right and Panel 3 left: injected data after filtering. We can clearly see the shifting lines in the residuals. This is to be compared with the filtered models during retrieval. Panel 3 right: vertical slices of panel 2 right at selected wavelengths. Each wavelength channel is a noisy light curve with continuum 1, where the deeper transit depth are from the absorption features. Panel 4 left: example of a noiseless, idealistic data consisting of stellar flux and shifting planetary flux only. Panel 4 right: injected models after filtering. This is to be compared with the residual data during retrieval. Panel 5 left: the stellar flux and shifting planetary flux of the noiseless data shown in panel 4 left. Panel 5 right: unit-less visualization of data and model at different stages. The similarity between filtered models and filtered data shows the effectiveness of our filtering process.}
\label{fig: injection_data_stage}
\end{figure*}

\newpage

\subsection{Data Detrending}
\label{sec: detrending}

In ground-based high-resolution data, the planet signal is hidden under a combination of instrumental noise, telluric noise, and stellar flux. In addition, the low planet-to-star contrast ratio makes signal extraction a challenge. However, one can leverage the strong Doppler variation of the planetary signal, coupled with data detrending techniques, to isolate the shifting planetary features from the background (assumed to be relatively stationary). Data detrending techniques applied to HRCCS commonly include Principal Component Analysis (PCA) \citep[e.g.,][]{Birkby2013,Birkby2018}, SYSREM \citep[e.g.,][]{Gibson2020,Gibson2022}, \correction{and several other techniques we discuss below}. High-resolution emission spectroscopy works as long as planets are visible in their orbits, unlike traditional low-resolution emission spectroscopy, which requires observations inside and outside an eclipse. This allows us to probe the atmospheres of the population of non-eclipsing exoplanets. High-resolution transmission spectroscopy, however, still relies on planets passing in front of their stars to produce an absorption signal. Since planetary emission is roughly proportional to $T^4$, while the transit depth is proportional to $(R_p$/$R_*)^2$, HRCCS favors hot, giant planets orbiting small stars.


Detrending data degrades the underlying planet signals \citep{Brogi2019,Gibson2020,Gibson2022}. This degradation is less important for the classical cross-correlation approach, where we can detect species as long as line positions of the template match with the residual data. For the purpose of retrievals, however, we can't obtain accurate constraints unless the forward model is a good representation of the residual data. This means that we must preprocess the forward model to mimic the degradation done on planet signal by data detrending. \cite{Brogi2019} injects the forward model into a representation of the background and applies the same filtering (PCA) to obtain a distorted version of forward model. This is proven to work well in \cite{Line2021}, \cite{Pelletier2021}, \cite{Pelletier2023}. However, this needs to be done on every forward model. As hundreds of thousands of forward models are needed for retrievals, it becomes a larger computational bottleneck than computing the high-resolution forward model. \cite{Gibson2020} applies SYSREM \citep{Tamuz2005} to filter the data and uses a high-pass filter to preprocess the model spectra. \cite{Gibson2022} introduces a fast-filtering approach that builds a linear basis for the background with SYSREM and preprosses the forward model by subtracting a linear fit to the basis. Since the linear least square fit has a closed-form solution, the time of this preprocessing is negligibly fast. We provide a mathematical description of the data detrending and model preprocesssing in Section \ref{sec: preprocess forward model}. We implement all the detrending methods in our framework, catering the needs of different applications.

\correction{
Other telluric correction routines such as molecfit \citep{Smette2015} that rely on synthetic telluric spectra may also be used (e.g. \citealt{Kesseli2022}). Recently, \citet{Maguire2024} tested both SYSREM and MOLECFIT for telluric removal on atmospheric retrievals of high-resolution optical transmission spectra, concluding there is no strong reason to favor either approach.
}

We implement PCA with Truncated Singular Value Decomposition for computational efficiency. \correction{The number of principal components to remove is usually determined empirically. Common approaches include optimizing against the real signal (CCF$_{\text{obs}}$), optimizing against an injected signal (CCF$_{\text{inj}}$), or optimizing against the difference between the two ($\Delta$CCF) \citep{Holmberg2022}. \cite{Cheverall2023} finds that optimizing $\Delta$CCF is more robust against over-optimization of noise and spurious signals. We test the three different approaches in Section \ref{sec: WASP77}.} The truncated SVD is applied directly to each individual order (two dimensional time-by-wavelength array) and is effectively picking out the first N eigenvalues that correspond to the dominant trend in time (background). We subtract the PCA-representation from original data to obtain the residual data (to be compared with preprocessed forward model). Figure~\ref{fig: injection_data_stage} demonstrates the effect of filtering with PCA.

As an improvement to PCA, SYSREM\footnote{An implementation of SYSREM can be found \href{https://github.com/stephtdouglas/PySysRem}{here}} removes systematic effects in a large set of light curve, taking heteroscedastic noise into account. We can view our data (for each order) as a collection of light curves, where the number of light curves equals the number of wavelength channels. SYSREM repeatedly finds the best rank-one approximation and subtracts it from the data. Each rank-one approximation is the outer product of two vectors, which can be interpreted as a basis vector and a weight vector. SYSREM therefore constructs a basis (not necessarily orthogonal) while filtering the background. SYSREM has the additional advantage of considering the heteroscedastic noise. In the limit of uniform noise, SYSREM is equivalent to PCA \citep{Tamuz2005}.

We also propose an alternative to build the matrix basis using Non-Negative Matrix Factorization (NMF) for the fast-filtering method. We use the implementation of NMF by \cite{Zhu2016}. Similar to SYSREM, NMF also considers heteroscedastic noise and can be built iteratively. Unlike SYSREM or PCA that picks out the dominant components, NMF best decomposes a matrix into additive non-negative components. \cite{Ren2018} proves the stability of NMF components and the linearity of NMF modeling when the components are constructed sequentially, rationalizing that preprocessing forward models can mimic distortion on signals. NMF is also more interpretable considering the non-negative nature of flux. We should note that NMF is good and picking out additive, stationary signals and should only be preferred over PCA for signals with negligible Doppler Shift. This primarily occur for planets in face-on orbits. NMF is used in direct imaging for removing the point-spread-function as an improvement to PCA \citep{Soummer2012}, \citep{Ren2018}. We simply point out the connection between the two applications of NMF and recommend the use of PCA/SYSREM over NMF for the aforementioned reasons.

\subsection{Time and Wavelength Dependent \\ Uncertainty Estimates}
\label{sec: uncertainties}
Due to the nature of telluric and instrumental noise, the uncertainties in flux level should be both time-dependent and wavelength-dependent. It is therefore important to have good estimates of the heteroscedastic noise \citep{Gibson2020}. We determine the uncertainties in flux values after the spectra are blaze corrected (see Section \ref{sec: observations}). We estimate the time and wavelength-dependent noise by assuming the noise is dominated by a Poisson term and a single background value, which takes the form $\sigma_i = \sqrt{a F_i + b}$, where $F_i$ is the measured flux for a given time and wavelength, and $a, b$ are coefficients to be determined. We subtracted a PCA model, preferably with a similar order to the number of SYSREM iterations, from the data to get a residual array $R_i$, which approximately leaves a zero-mean array plus noise. The coefficients a and b are found by fitting our noise model (i.e. $\sigma_i$ above) to each order, by optimizing the likelihood (after removing constant terms):
\[ \ln \mathcal{L}(a, b) = -0.5 \displaystyle\sum_i \left(\frac{R_i}{\sigma_i} \right)^2 - \displaystyle\sum_i \ln \sigma_i\]

Finally, we reconstructed the uncertainties array using the best-fitting values for a and b. As this array is in turn noisy, we then modelled it with another 5th-order PCA model, and this model becomes our final estimate of the noise. This final step is required to remove any bias in the noise determination. This is due to a subtle effect caused by low values of $F_i$ always having lower uncertainties due to the assumption of Poisson noise, whereas in reality there is noise in the determination of the uncertainties. This can bias the ﬁts when modelling residual data sets. See \cite{Gibson2022} for further details, as well as a visualization of the noise determination.

\subsection{Preprocessing the Forward Model}
\label{sec: preprocess forward model}
The intrinsic rotation of a planet can have broadening effect on the lines. The planetary rotational broadening effect is related to the projected planetary rotational velocity ($V \sin i$), where $V$ is the planet's rotational velocity and $i$ is the inclination of the planet's rotation axis \citep{Spring2022}. The resolution of the instrument also limits how refined the lines we observe are and effectively broadens the observed lines as well. We have two options to account for broadening: (i) convolve with a rotational broadening kernel and an instrumental broadening kernel with assumed widths similar to \cite{Line2021}; or (ii) choose to retrieve for the broadening effect by adding an ad hoc parameter prescribing the width of the broadening kernel \citep{Gibson2022}.

After accounting for line broadening, we can shift the model to the data wavelength grid (accounting for the appropriate Doppler shift) for each order. This is typically done by by interpolation (e.g. linear interpolation, spline interpolation, etc.).

The data are shifted to stellar rest frame during preprocessing stage, which includes correction for barycentric velocity and systemic velocity. In stellar rest frame, the planet's velocity $V_p(t)$ is
\[ V_p(t) = K_p \sin(2\pi*\phi(t)) + V_{sys} \numberthis \label{eq: planet velocity} \]
where \correction{$\phi(t)$ is the orbital phase of the exposure at time $t$}, $K_p$ is the semi-amplitude of the planet radial velocit,  $V_{sys}$ is is a correction term that should be centered at 0 as a measure of any additional Doppler shifting (since systemic velocity is already corrected). $V_{sys}$ is sometimes referenced as $\Delta V_{sys}$ to avoid ambiguity.

For each pair of $V_{sys}$ and $K_p$ in the non-relativistic regime, we can Doppler-shift the model spectrum $m$ with
\[ \lambda' = \lambda(1 - \frac{V_p}{c})\]
\[ m'(\lambda') = m(\lambda) \]
or equivalently,
\[ \lambda' = \lambda(1 + \frac{V_p}{c})\]
\[ m'(\lambda) = m(\lambda') \]
where $m'$ is the shifted model to be compared with data, $\lambda$ and $\lambda'$ are the wavelengths in planet's rest frame and observer's rest frame respectively.

Finally, we preprocess the forward model to account for the signal degradation by data detrending. At this stage, we have a two-dimensional matrix (exposures by pixels) for each order. This process depends on how you detrend data as well as the type of spectroscopy.

For PCA on emission spectroscopy \citep{Brogi2019}, we remove the PCA model from the data matrix. We divide the forward model by a stellar model spectrum (a PHOENIX model) to obtain $F_p/F_s$, which provides us an estimation of planet-to-star flux ratio. We save the PCA model, interpreting it as the telluric-systematic-stellar background. We inject the model into the PCA model and apply PCA again to preprocess the model. This process can be seen as
\begin{align*}
    F_{\text{injected}} &= (1+\frac{F_p}{F_s}) (C_s + C_{\text{noise}}) \\
    &=  C_s \frac{F_p}{F_s} + C_s + C_{\text{noise}} + \frac{F_p C_{\text{noise}}}{F_s} \\
    &\approx C_p + C_s + C_{\text{noise}} \numberthis \label{eq: emission filter}
\end{align*}
where $C$ stands for detector counts and $F$ stands for flux. Notice the last term in the second line is negligible compared to others and can be omitted. We use the fact that $F_s / C_s  \approx F_p / C_p$ (i.e. detector gain is uniform). We notice the injection places the model into the same background as the actual planet signal, and re-applying PCA effectively reproduces the same distortion on the model spectrum. Note that we can't simply add $F_p$ into $C_s + C_{\text{noise}}$ because the detector counts doesn't reflect the true flux level.

For now we do not consider applying PCA to transmission spectroscopy, as it is slower and ignores the time-wavelength dependency of noise compared the fast-filtering approach proposed in \cite{Gibson2022}. We note that this approach also work for emission spectroscopy, which has been achieved in \citet{Ramkumar2023}. We discuss fast-filtering for both emission and transmission spectroscopy next.

The key of the fast-filtering approach is finding a low-rank basis that best approximates the data matrix. As we note earlier in Section \ref{sec: detrending}, PCA, SYSREM, NMF all achieve this purpose. We denote the the set of basis vectors as $U$. We interpret the residual data (after subtraction of the best low-rank approximation) as the distorted planet signal. Be aware that the fast-filtering process differ slightly depending on the spectrum type (emission or transmission).

The left column in Figure~\ref{fig: injection_data_stage} is a step-by-step visualization of fast-filtering for emission spectroscopy. We construct a set of basis vectors $U$ and a set of weights $W$ that best approximate the raw flux $D$ for each order. Denoting the residual data as $R$, this process is 
\[ D = U W + R \]
Interpreting $U W$ as the telluric-systematic-stellar background, we multiply $F_p/F_s$ into $U W$ and fit the resulting matrix to the basis $U$ before subtracting the fit from the data. This is similar to PCA on emission spectroscopy (Equation \ref{eq: emission filter}).

The right column in Figure~\ref{fig: injection_data_stage} is a step-by-step visualization of fast-filtering for transmission spectroscopy. We first normalize the data to have continuum 1 by dividing out the median out-of-transit spectrum for each order. The data now represent the relative ratio compared to the out-of-transit spectrum. Each wavelength channel becomes a (noisy) transit light curve, where transit depths are higher where the absorption lines are (see panel 4 of Figure~\ref{fig: injection_data_stage}). This is to be compared with our transmission forward model later. The uncertainties for each order determined earlier are divided through by the median spectrum to account for the preprocessing (subtraction of the SYSREM model does not modify the uncertainties). 

\begin{figure}[t]
\centering
\includegraphics[width=0.475\textwidth]{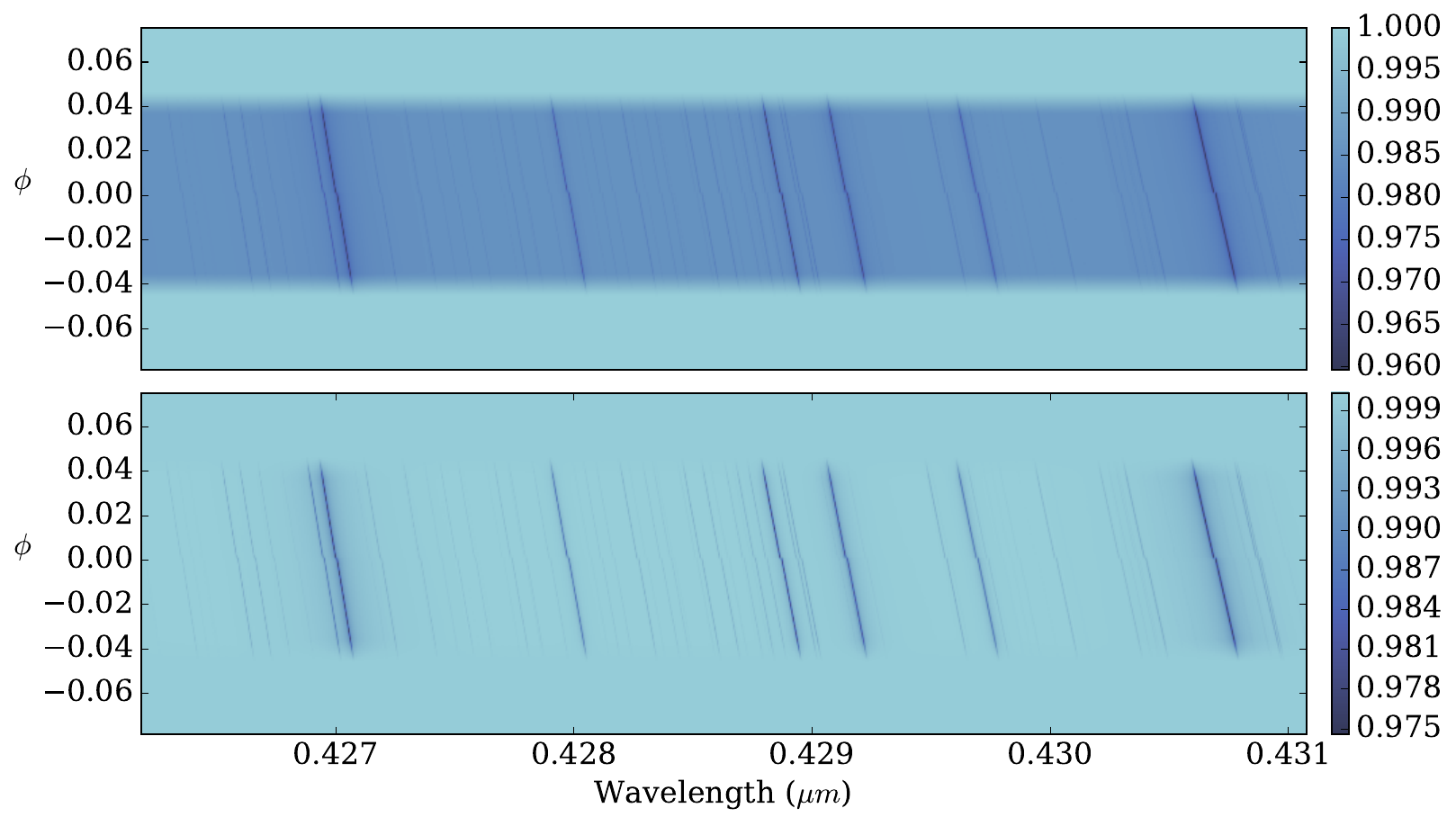}
\caption{Upper panel: Example transmission spectra shifted according to planet velocities at each phase weighted by the transit shape. Notice out-of-transit spectra are equal to 1. Lower panel: Removing the transit shape by dividing the median of wavelength for each phase.}
\label{fig: transmission_shifted}
\end{figure}

We then weight the model spectrum according to the transit shape and add 1 to each spectrum for normalization. The process is
\[M = \frac{1-w(t)}{\max(1-w)} (-F_p) + 1 \numberthis \label{eq: model shifting}\] 
where $F_p$ is the transmission spectrum in transit depth $R_p^2/R_*^2$ and $w(t)$ is the phase-dependent transit shape. We compute the transit shape with the Batman package \citep{Kreidberg2015}. We also divide the median of wavelength for each phase. This is to mimic the effect of blaze correction on the data (see Section \ref{sec: observations}), which brings the time-series of data to a common flux level. These steps ensure the model have the same form as the signal. Figure~\ref{fig: transmission_shifted} is a demonstration of the transmission forward model before preprocessing. We then fit this matrix to the basis $U$.

The fitting process for both emission and transmission mimics the effect of detrending on the planet's signal. We default to the linear least squares fit for computational efficiency as it has a closed form solution,
\[ W' = (U^TU)^{-1} U^T Y = U^\dagger Y\]
for a 2D (time-wavelength) array Y, where the term $(U^T U)^{-1} U^T$ is the Moore-Penrose inverse, $\rm U^\dagger$. The best-fit model, Y', to the array Y, is then given by the outer product of the best-fit weights, W', and the basis vectors U,
\[Y' = U W = U U^\dagger Y\]

For transmission spectroscopy, we introduce a bias term (a vector filled with ones) to the basis vectors U. The purpose of the extra basis vector is to account for the potential offset induced from dividing through by the median spectrum in each order (the median spectrum might not accurately reflect the out-of-transit baseline) \citep{Gibson2022}. To account for the heteroskedastic data uncertainties, which were accounted for initially when computing U, we take the mean of the uncertainties over wavelength $\hat \sigma$ for each order, as it still captures the major time-dependent trends for each wavelength channel and pertains the closed-form solution. 

Therefore, the final fit to the forward model is:
\[ M' = U(\Lambda U)^\dagger (\Lambda M ) \numberthis \label{eq: model fit}\]
where $\Lambda$ is a diagonal matrix with $1/\hat \sigma$. $M-M'$ is the final model to be compared with the residual $R$ during retrievals. Notice $U(\Lambda U)^\dagger \Lambda$ does not depend on the model and therefore can be stored to further reduce the computational cost \cite{Gibson2022}.

\subsection{Likelihood Mapping and Retrieval}
\label{sec: likelihood}
We acknowledge that the following derivation has appeared several times in the literature (e.g. \citealt{Brogi2019, Gibson2020}). We include a brief discussion of the likelihood mapping for completeness.

We start by assuming that the noise is Gaussian distributed at each pixel $i$ with standard deviation $\beta \sigma_i$, where $\sigma_i$ is the uncertainties at pixel $i$ and $\beta$ is the scale constant for noise \citep{Gibson2020, Gibson2022}. Then, the full Gaussian likelihood is 
\[ \mathcal L(\theta) = \prod_i \frac{1}{\sqrt{2 \pi (\beta \sigma_i)^2}} \exp \left ( -\frac{1}{2}\frac{(f_i - \alpha m_i)^2}{(\beta \sigma_i)^2} \right) \]
\[ \ln \mathcal L(\theta) = -\frac{1}{2}\chi_\beta^2 - N\ln \beta - \sum_i \ln \sigma_i - \frac{N}{2} \ln 2\pi \]
where $ \chi_\beta^2$ is defined as
\begin{align*}
    \chi_\beta^2 & = \sum_i \frac{(f_i - \alpha m_i)^2}{(\beta \sigma_i)^2} \\
&= \frac{\chi^2}{\beta^2} \\
\chi^2 &= \sum_i \frac{f_i^2}{\sigma_i^2} + \alpha^2 \sum_i \frac{m_i^2}{\sigma_i^2} - 2\alpha \text{CCF} \\
\text{CCF} &= \sum_i \frac{f_i m_i}{\sigma_i^2}
\end{align*}

The last two terms are constant and can be dropped during parameter estimation, leaving
\[ \ln \mathcal L(\theta) = -\frac{1}{2}\chi_\beta^2 - N\ln \beta \numberthis \label{eq: full logL}\]

Following the same nulling procedure described in \cite{Brogi2019}, we can null the dependence on $\beta$ by maximizing $\ln \mathcal L(\theta)$ with respect to $\beta$ by solving
\begin{align*}
    \frac{\partial \ln \mathcal L(\theta)} {\partial \beta} \rvert_{\hat \beta} &= 0 \\
    \hat \beta^2 = \frac{\chi^2}{N}
\end{align*}

Using the maximum likelihood estimator (MLE) $\hat \beta$ and dropping the constant terms, we get
\begin{align*}
   \ln \mathcal L(\theta) &= - \frac{N}{2} \ln \frac{\chi^2}{N} \numberthis \label{eq: nulled b logL}
\end{align*}

We can therefore exclude $\beta$ in retrievals and instead use Equation \ref{eq: nulled b logL}. By doing so, we are effectively optimizing $\beta$ for each order if we calculate likelihood per order. If we stack all the orders and calculate the likelihood with Equation \ref{eq: nulled b logL}, we are optimizing $\beta$ for the entire dataset. However, we have defined $\beta$ to be the scaling of the uncertainties, which should be a property of the data only, whereas $\hat \beta$ also depends on the model. \citet{Gibson2020} and our own experiments both show very little difference between these choices. \citet{Klein2024} also discusses this subtlety and report similar findings. Therefore, we provide options to stack the orders and not fit for $\beta$, which improves the efficiency of retrievals. 

If we further assume the uncertainty $\sigma$ is the same at each pixel, we will recover equation (8) in \citet{Brogi2019}. However, we recommend to fit for uncertainties considering the heteroscedastic nature of detector noise as described in Section \ref{sec: uncertainties}, considering it does not cause extra overhead for retrievals.

With the likelihood mapping defined, we can finally perform retrievals. In addition to parameters for atmospheres, we include the aforementioned parameters specific to high-resolution retrievals: $\alpha, \beta, K_p, V_{sys}, W_{conv}$. Here $\alpha$ is a scaling factor for model, $\beta$ is a scaling factor for uncertainties, $W_{conv}$ is the width of broadening kernel. We feed the parameterized model, the likelihood mapping, and the prior to \POSEIDON's sampler. \POSEIDON uses Pymultinest \citep{Feroz2019}, a Python wrapper for nested sampling that efficiently samples the parameter space, estimates Bayesian evidence, and provides posterior distribution of the parameters.

\subsection{Classical Cross-correlation Approach}
\label{sec: CCF map}
Before \cite{Brogi2019} introduced a robust log-likelihood mapping and a retrieval framework, HRCCS had already been and still is an excellent tool for detecting chemical species. We implement the classical cross-correlation in our framework and describe the process below.

We cross-correlate a template model spectrum with the data. If the model is a good representation of the true planet signal, we expect a peak cross-correlation function (CCF) at the correct radial velocity shift \citep{Birkby2018}. This is only sensitive to the location of lines and much less sensitive to the shape of lines. The distortion of signal is therefore less important for the classical cross-correlation approach. Modeling this effect, however, should still be able to boost the detection significance \citep{Gibson2022}.


For each phase and each order, we cross correlate the template model with the the extracted exoplanet signal that survives the filtering with a velocity range given by prior information of the Keplerian planet velocity $K_p$ and systemic velocity $V_{sys}$. \correction{We typically choose a velocity range at least $\pm 100$\,km\,h$^{-1}$ the published values to avoid the peak being submerged by neighboring values. The actual velocity ranges used in the paper can be seen on the axes of plots.} The planet radial velocity is related to these two quantities by Equation \ref{eq: planet velocity}. The template is then Doppler-shifted by a range of planet velocities using linear interpolation (Figure~\ref{fig: transmission_shifted}).

The cross-correlation function (CCF) is then obtained by multiplying the shifted model with the data and integrating over wavelength
\[ \text{CCF}(V_p) = \sum_i \frac{f_i m_i (V_p)}{\sigma_i^2}\]
and therefore can be thought as a function of the planet velocity $V_p$. Here, $f_i$, $m_i$, $\sigma_i$ are the (residual) data, model spectrum, and uncertainties at each pixel.

This will produce a 1D array of CCF v.s. radial velocity for each phase and each order. Integrating over order, we get a 2D matrix of CCF, where the two dimensions are phase and radial velocities (e.g. Figure~\ref{fig: WASP77_CCF}). We expect the peak values at each phase forms a trail with slope approximately equal to $1/K_p$ \citep{Birkby2018}, since $\Delta V_p = K_p \sin \Delta \phi \approx K_p \Delta \phi$ for small phase during transit. We can integrate the results over all phases to fully utilize the cross-correlation results. To do so, we shift the CCF matrix to planet rest frame assuming $K_p$ is known. We then sum the CCF values over time and obtain a phase-folded array of CCF v.s. radial velocity. Since we do not know $K_p$ perfectly, we repeat the same process for a range of possible values of $K_p$ and obtain a $K_p$-$V_{sys}$ map of CCF, where we expect a peak value at the true $K_p$-$V_{sys}$ (e.g. Figure~\ref{fig: WASP77_CCF}).

\begin{figure}[t]
\centering
\includegraphics[trim=20 0 -10 0, clip, width=0.49\textwidth]{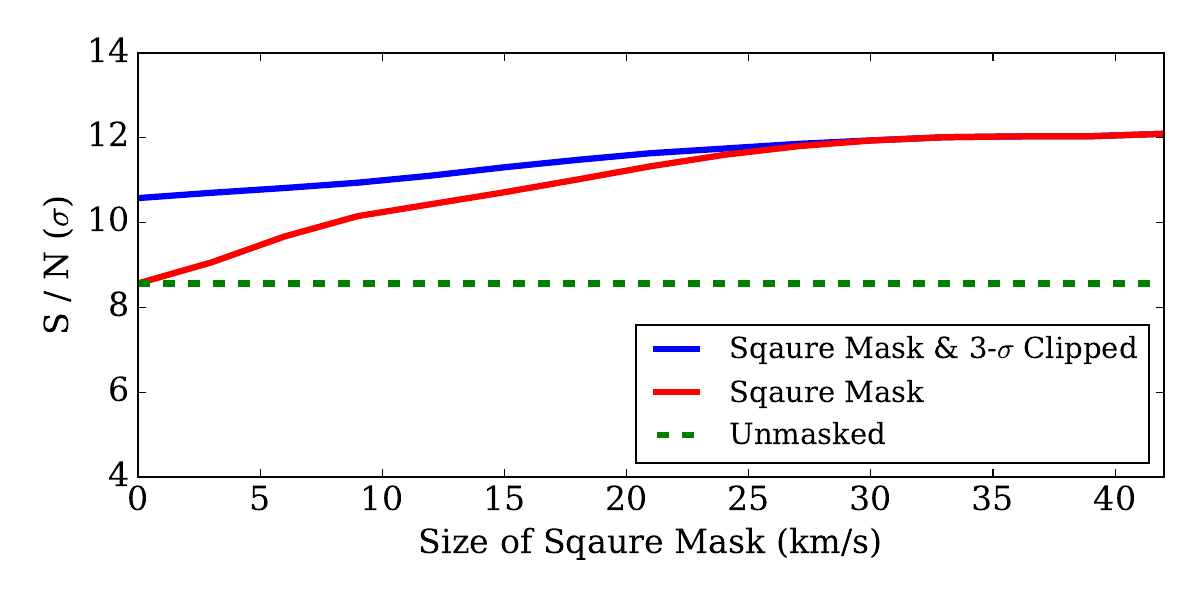}
\caption{\correction{Detection significance (S/N) versus the size of the square mask used when calculating the standard deviation of the CCF map. The standard deviation is calculated after clipping values outside $3 \sigma$ for the blue curve, in addition to applying the square mask centered at the expected velocity.}}
\label{fig: CCF_mask}
\end{figure}

We calculate the detection significance by taking the ratio between the peak value of the total cross-correlation and the standard deviation of the cross-correlation coefficients. When calculating the standard deviation in this study, we mask the region near the peak in velocity space ($\pm 20$\,km\,h$^{-1}$). This has been the common practice to exclude the measured signal for its simplicity, but is criticized for its dependence on masked region. \correction{Figure~\ref{fig: CCF_mask} demonstrates how the detection significance changes as we change the masked region either with or without sigma clipping. Clearly, a large masks artificially reduce the standard deviation and can lead to overly-optimistic detection. As a rule of thumb, we suggest using the minimum size of a square mask that includes 95\% of outliers as determined by sigma clipping (i.e. where the blue and red curves start to overlap in Figure~\ref{fig: CCF_mask}).} Detection significance can also be measured with alternative methods such as the Welch t-test \citep{Welch1947, Birkby2017, Cabot2019} and phase-shuffling \citep{Esteves2017, Merritt2020, Merritt2021}.

The center-to-limb variation (CLV) and the Rossiter–McLaughlin (RM) effect could be corrected to improve detection. Both are due to spatial variations of the stellar spectrum across the stellar disc, causing small misalignment in the stellar spectra. Their effects on cross-correlation, termed "Doppler Shadow", are demonstrated in \citep{Maguire2023}. See \citet{Turner2020} and \citep{Nugroho2020} for discussions on how to remove RM and CLV.


\section{High-Resolution Observations of WASP-77Ab and WASP-121b}
\label{sec: observations}

\subsection{WASP-77Ab IGRINS Emission Spectrum}
\label{subsec:WASP-77Ab_observations}

\begin{figure}[ht]
\centering
\includegraphics[width=0.465\textwidth]{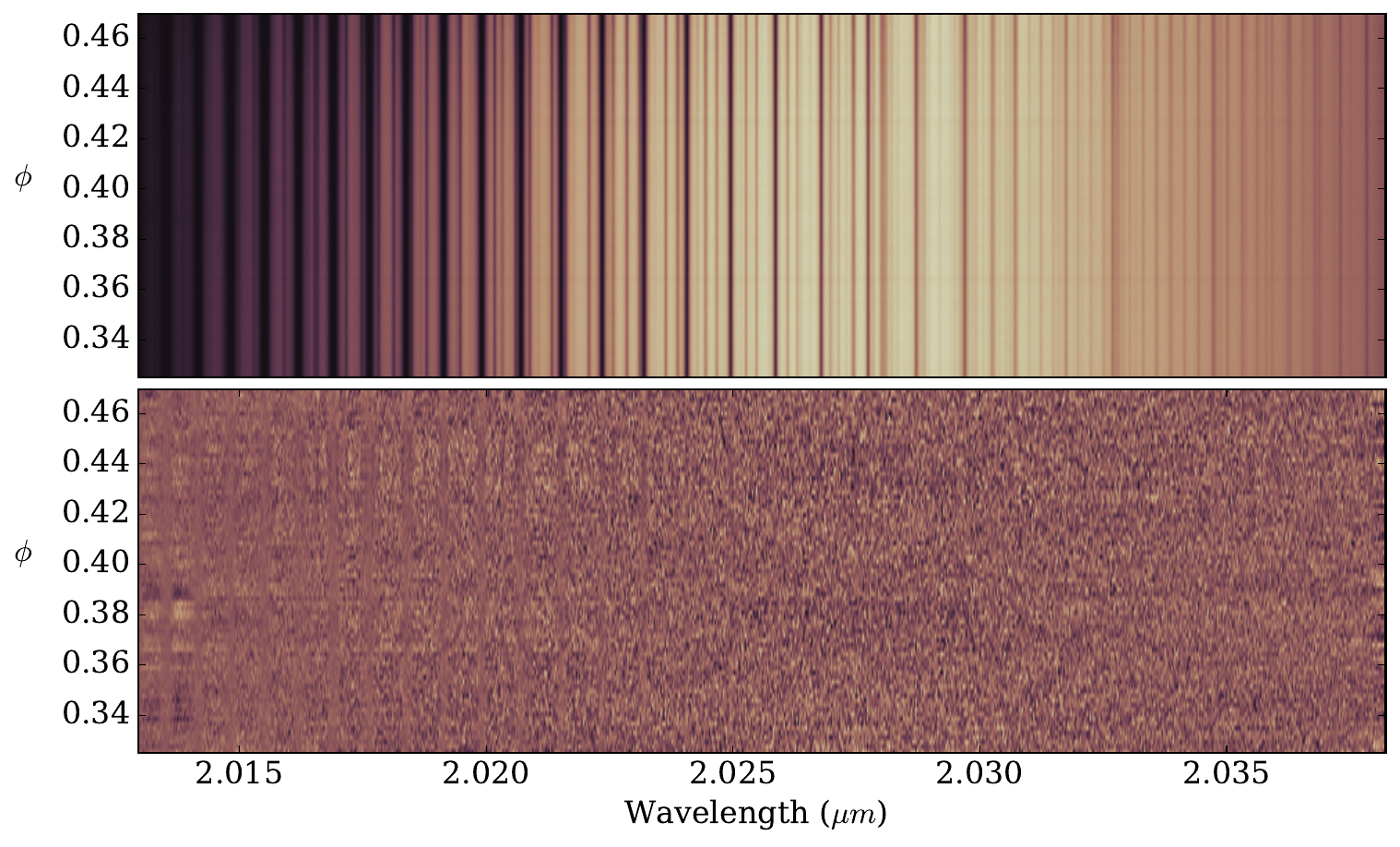}
\caption{Upper panel: the reduced IGRINS observation of WASP-77Ab. The flux at each phase is calibrated to the same level. The emission signal of WASP-77Ab is buried under the background and noise. Lower panel: Residual after removing the first 4 principal components. The residual should be a good representation of the signal.}
\label{fig: WASP77_data}

\includegraphics[width=0.475\textwidth]{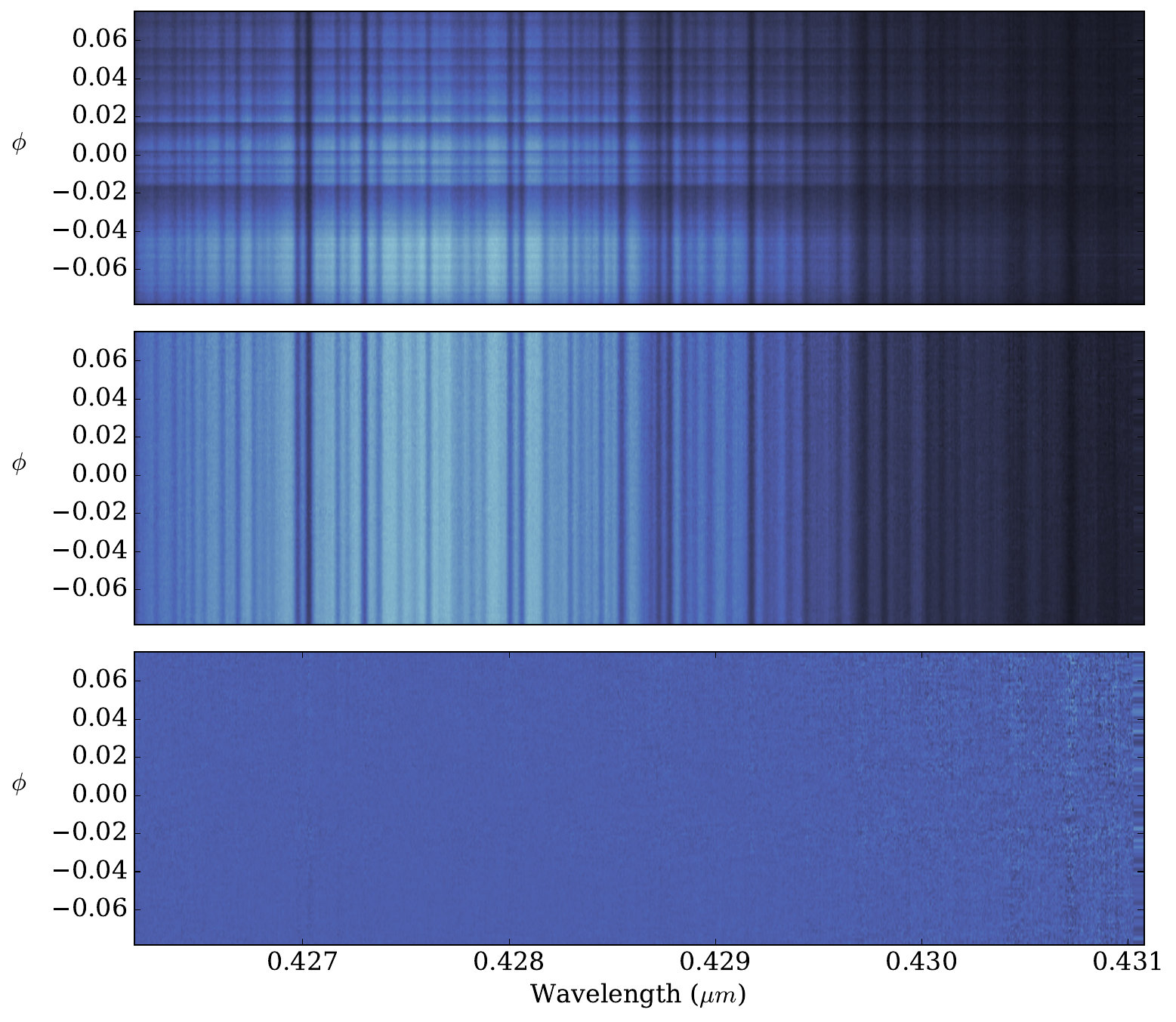}
\caption{Upper panel: the raw UVES observation of WASP-121b. Middle Panel: the observation after blaze correction. The transmission features of the planet is buried under the background and noise. Lower panel: residual after filtering.}
\label{fig: WASP121_data}
\end{figure}

The IGRINS pre-eclipse observation of WASP-77Ab, from December 14, 2020, is described and first analyzed in \citet{Line2021}. This observation spans a 4.7\,hr continuous time-series with 79 exposures (140 seconds per exposure) covering phases, $\phi$, from 0.32 to 0.47. The spectral range covers 1.43--2.42\,$\micron$ with a resolution of R $\sim$ 45,000 over 54 spectral orders. Figure~\ref{fig: WASP77_data} provides a visualization of a single order from the observation. We remove bad pixels via a fit to the raw data and then replace outliers with the fitted model. We find that outlier removal does not bias our analysis results, since less than 0.1\% of pixel values are marked as outliers for each dataset.

\subsection{WASP-121b UVES Transmission Spectrum}
\label{subsec:WASP-121b_observations}

The UVES transit observation of WASP-121b \citep{Gibson2020} is analyzed in \cite{Merritt2021} and \citet{Gibson2022}. It consists of a single transit observed on December 25, 2016, as part of program 098.C-0547 (PI: N. Gibson). The observations have an exposure time of 100 s (with a readout time of 24 s), and 137 exposures over 4.7\,hr with approximately 83 exposures during transit. Exposures 70–72 were clipped from the data analysis due to loss of guiding. The observations were separated into red and blue spectral arms. The blue arm covers 0.37--0.50\,$\micron$ over 31 spectral orders with a resolution of R $\sim$ 80,000. The red arm covers 0.49--0.87\,$\micron$ over 54 spectral orders with R $\sim$ 110,000. Figure~\ref{fig: WASP121_data} shows a single order of this observation. We remove bad pixels as described above for WASP-77Ab.

Next, we apply a blaze correction following the procedure outlined in \cite{Gibson2020}. The procedure we apply does not remove the blaze function, but places every spectrum (for each order) on a ‘common’ blaze. This accounts for time-dependent distortions of the instrument response, and improves the performance of the filtering methods. To achieve this, we divide each spectrum through by the median spectrum in each order, smooth each residual spectrum with a median and then Gaussian filter to get a blaze correction relative to the median spectrum, and finally divide each of the original spectra through by its respective blaze correction. The width of the filters are determined empirically. \cite{Gibson2022} have chosen 501 and 100 as the widths for the median and Gaussian filters and reported better performance than 50 and 11 as the widths in \cite{Gibson2020}. Small drifts in wavelength solutions can be remedied by aligning data with a template, e.g. a stellar model or a telluric model, through cross-correlation. After wavelength alignments, the data should be in the rest frame of the star and we apply the fast-filtering method as described in Section \ref{sec: detrending} and \ref{sec: preprocess forward model}. Figure~\ref{fig: injection_data_stage} is a demonstration of the filtering process on simulated data. We can see the time-dependent absorption lines are preserved after the filtering, whereas the stationary features are removed.

\section{Injection Tests}
\label{sec: injection}
\subsection{Generating Simulated Data}
To validate the methods, we run a series of tests on simulated data injected by a known signal. To keep the test as realistic as possible, we inject the signal directly into real data, in our case the UVES observation of WASP-121b for transmission and IGRINS observation of WASP-77Ab for emission as described above. The ensures the noise level of the injection tests to match what's expected in real-life applications. To avoid mixing with the real exoplanet signal, we choose $K_p=-200$\,km\,s$^{-1}$ and $V_{sys}=-20$\,km\,s$^{-1}$ that are opposite to the real signal in velocity space.

For both types of spectroscopy, we model the atmosphere as one dimensional with constant vertical abundances consisting of 100 layers from $10^{-5}$ bar to 100 bars. The P-T parametrization and true parameter values of each injection test can found in the Appendix. \correction{The abundances used here are meant to mimic typical abundances found in Hot Jupiters but can be arbitrarily chosen as long as they can produce spectral features. Exact values used to create the injected signals can be found in Table \ref{tab:injectiontests}.} To account for the scale factor $\alpha$ in the injection, we re-compute the continuum without the chemical species. We subtract the continuum from the full model, scale the difference by $\alpha$, and add the continuum model back to the model. Notice that, this is only done for generation the injection signal. In the case of retrievals, we can leave the continuum to be removed by the model filtering. We then shift the 1D model with an appropriate planet velocity for each order at each phase, resulting in a 3D (order-time-wavelength) model. To prepare the emission injected data, we multiply the shifted planet-to-star signal ($F_p/F_s$) and add into the raw data. To prepare the transmission injected data, we weight the shifted model by the transit shape of WASP-121b (Equation: \ref{eq: model shifting}) before multiplying into the raw data. The weighting ensures the out-of-transit model is one and in-transit model accurately reflect the transit depths at each phase and wavelength. We note that the signal injection can be done either before or after blaze correction for each order. We have found that blaze correction does not modify the underlying signal. We apply fast-filtering with SYSREM in the same fashion as described in Section \ref{sec: detrending} and \ref{sec: preprocess forward model}.

\subsection{Retrieval setting}
\label{sec: injection retrieval setting}
With injected data prepared, we are now ready to perform retrieval analysis using the methods outlined in Section \ref{sec: likelihood}. We parameterize the atmosphere in the same fashion as the injected signal. We include a test to evaluate validity of assuming atmospheres to be isothermal for transmission spectroscopy. In practice we want the retrieval to be as efficient as possible, so it is valuable to test whether the assumption of isothermal atmosphere introduces a strong bias. 

For each set of parameters, we compute the forward model, apply the model shifting and filtering following Section \ref{sec: preprocess forward model}. We ﬁt each 2D array (time versus wavelength) using the basis models (Equation \ref{eq: model fit}), and finally subtract the result from the model ﬁt. This is to be cross correlated with the data residuals and be converted into log-likelihood. \POSEIDON's retrieval framework, taking the log-likelihood function and the prior as input, samples the parameter space and outputs a posterior distribution of the parameters.

\subsection{Results}
Since high-resolution retrievals with PCA have been tested in previous studies \citep[e.g.,][]{Brogi2019, Gibson2022}, and our results applied to real datasets match their results well, our injection tests focus on testing SYSREM. We are interested in whether or not SYSREM can provide computational advantage while providing accurate constraints on these parameters.

The injected model contains four species \ce{H2O, CO, CH4, NH3} with $\log X_{\rm species}=-4$ with the same T-P profile. Our results demonstrate that SYSREM with HRS can retrieve information accurately for models containing up to four chemical species (Figure~\ref{fig: WASP77_injection_H2O_CO_CH4_NH3_combined}). Note that we fixed the planet radius for this injection test. As we will see in Figure~\ref{fig: WASP77_corner_combined}, there's a degeneracy between planet radius and the scale parameter. This nuance is less of an issue as we often have reliable radius measurements for objects of interests to HRS.

Our transmission injected model is parameterized by the six-parameter P-T \citep{Madhusudhan2009} and mixing ratios of three species \ce{Fe, Mg, Cr}, where $\log X_{\rm species}=-6$. We test the isothermal assumption here by comparing the results using the parameterized P-T model and the isothermal model. We conclude that transmission spectroscopy is not sensitive to temperature gradient even though the true signal has temperature variations, which justifies the isothermal assumption (see Figure~\ref{fig: WASP121_injection_all_retrieved}). In contrast, emission spectroscopy depends on temperature gradient to first order and therefore requires flexible T-P parameterizations. 

We should note the scaling parameter $\alpha$ is about half the injected value. This is not alarming since $\alpha$ serves as a nuance parameter compensating for the effect of filtering on the planet signal. The offset shown in retrieved scaling could as well be attributed to the difference between retrieved T-P profile and true profile. We confirm the scaling of the signal remains unchanged after the filtering through a noiseless realization of data, and we have run another experiment demonstrating the scaling can be correctly retrieved when the true profile is isothermal. Since the abundances are accurately constrained, we conclude that the framework is indeed reliable.

Additional parameters as well as priors for both tests can be found in Table \ref{tab:injectiontests}.

\section{Application: High-Resolution Emission Spectroscopy of WASP-77 Ab}
\label{sec: WASP77}
\subsection{Retrieval setting}
\label{sec: prior}
Since emission spectrum is highly dependent on temperature gradient, we parameterize the cloud-free atmosphere with the six-parameter temperature profile scheme \cite{Madhusudhan2009} and constant-with-altitude volume mixing ratios for gases considered (\ce{H2O, CO, CH4, NH3}). The atmosphere has 100 layers equally spaced in log space between 0.01 bar and 100 bar. We include the scale parameter $\log a$, Keplerian planet velocity $K_p$, and system velocity $V_{sys}$ as free parameters. We model the line broadening by convolving the model spectra with a rotational kernel followed by an instrumental kernel and fix the planet radius to reproduce the exact settings of \cite{Line2021}. We also run retrievals with the planet radius as a free parameter. Unless specified otherwise, we use the same prior for all parameters across the paper, except for planet-dependent ones, i.e. $K_p$, $V_{sys}$, $R_p$. For $K_p$ and $V_{sys}$, we use a uniform prior ($\pm 50$\,km\,h$^{-1}$) centered at the published velocities. For $R_p$, we use a Gaussian prior with width 0.05 $R_J$ centered at the published radius. Priors for the remaining parameters can be found in \ref{tab:injectiontests}.

\subsection{Results}
\label{sec: looser constraint}
\begin{figure}[t]
\centering
\includegraphics[trim=20 0 -10 0, clip, width=0.49\textwidth]{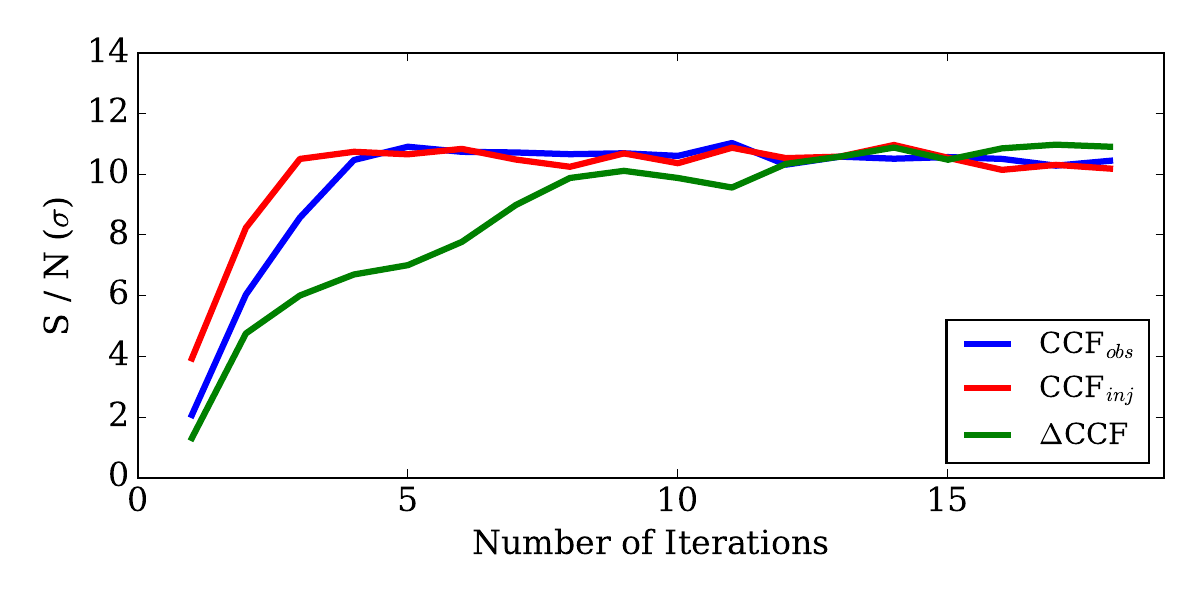}
\caption{\correction{Detection significance (S/N) versus number of PCA iterations. Unlike \cite{Cheverall2023}, we do not find that detection strengths decay as more principal components are removed, nor do we find a significant advantage for one approach over the others. We therefore use 10 iterations throughout our analysis. A data-driven approach, which halts the iterations when the variance of the residuals plateau, provides a similar number.}}
\label{fig: CCF_compare}
\end{figure}

\begin{figure*}[t]
\centering
\begin{subfigure}
    \centering
    \includegraphics[width=0.49\textwidth]{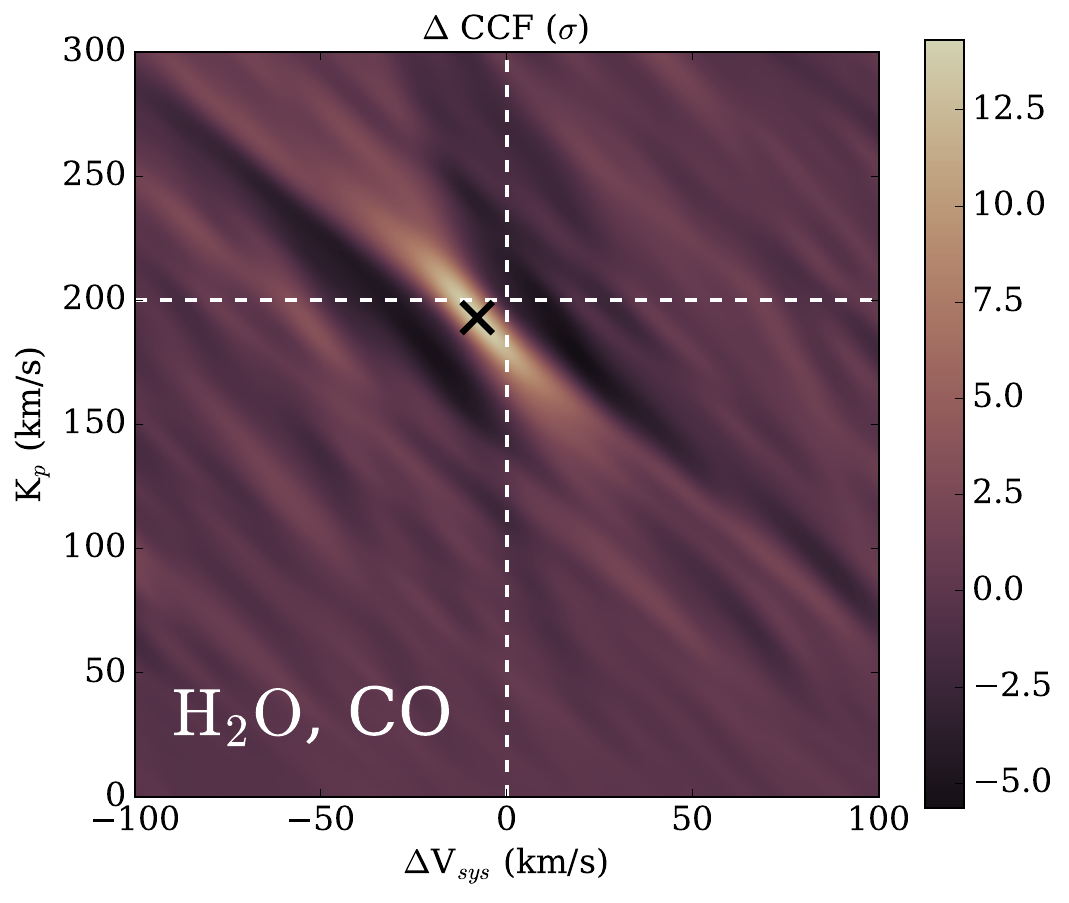}
\end{subfigure}
\begin{subfigure}
    \centering
    \includegraphics[width=0.49\textwidth]{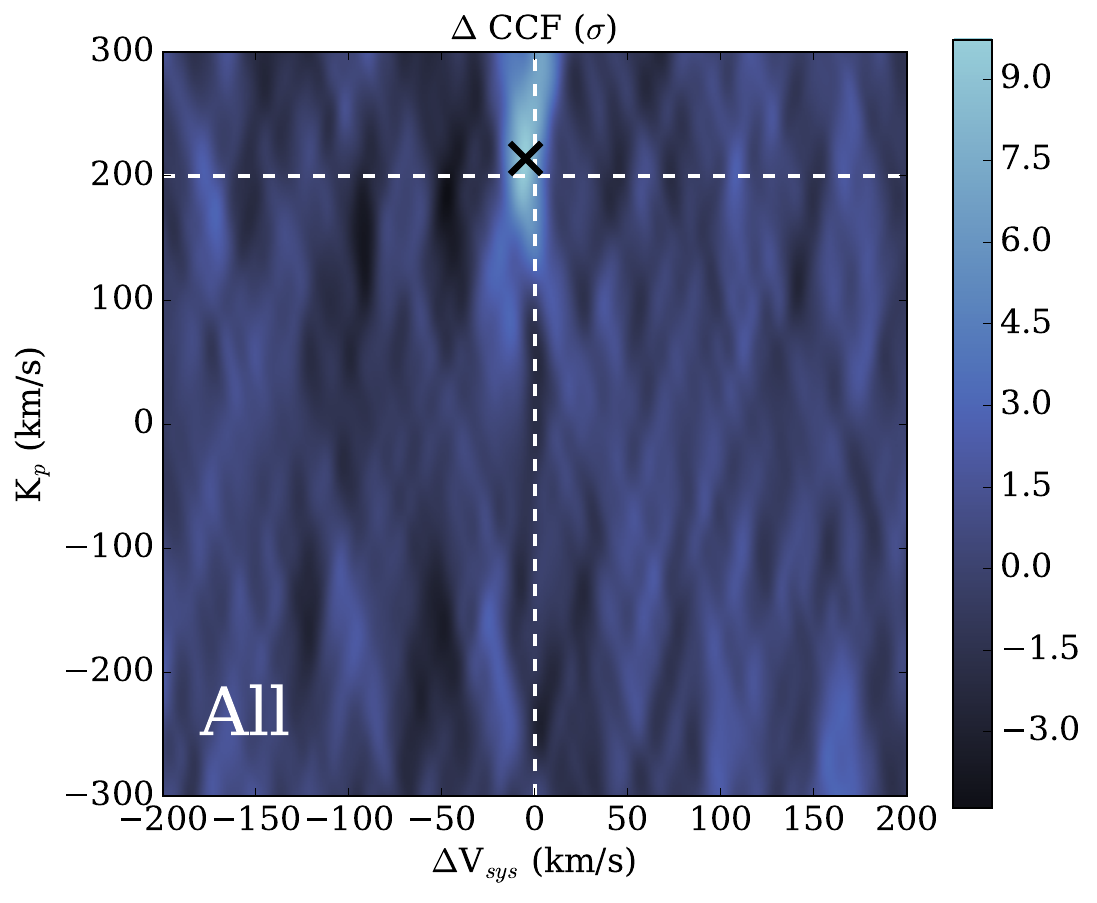}    
\end{subfigure}
\begin{subfigure}
    \centering
    \includegraphics[width=0.49\textwidth]{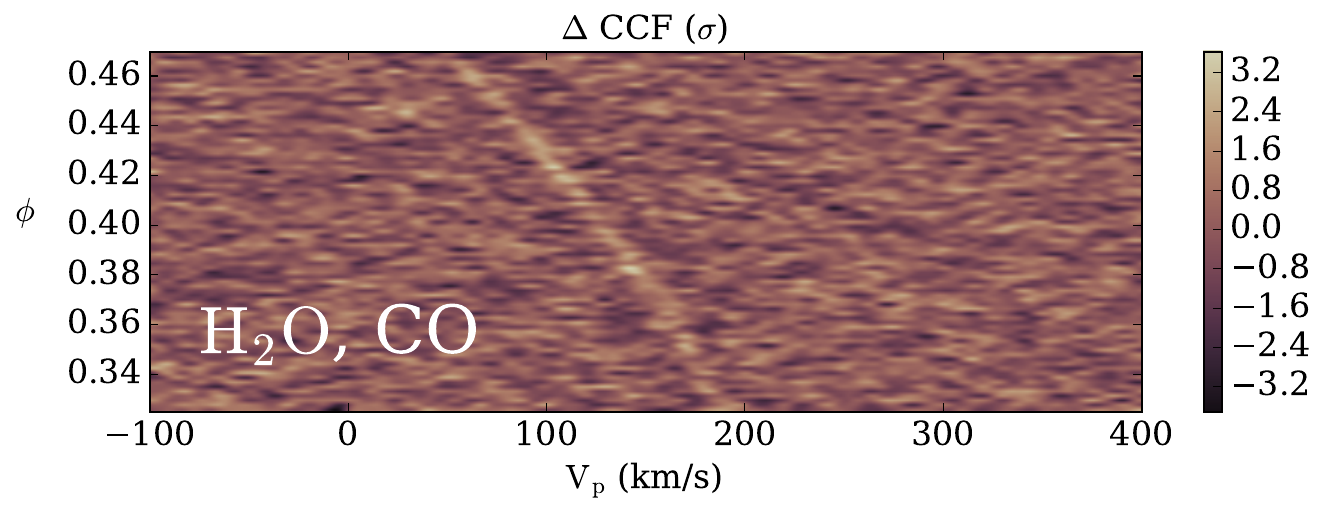}
\end{subfigure}
\begin{subfigure}
    \centering
    \includegraphics[width=0.49\textwidth]{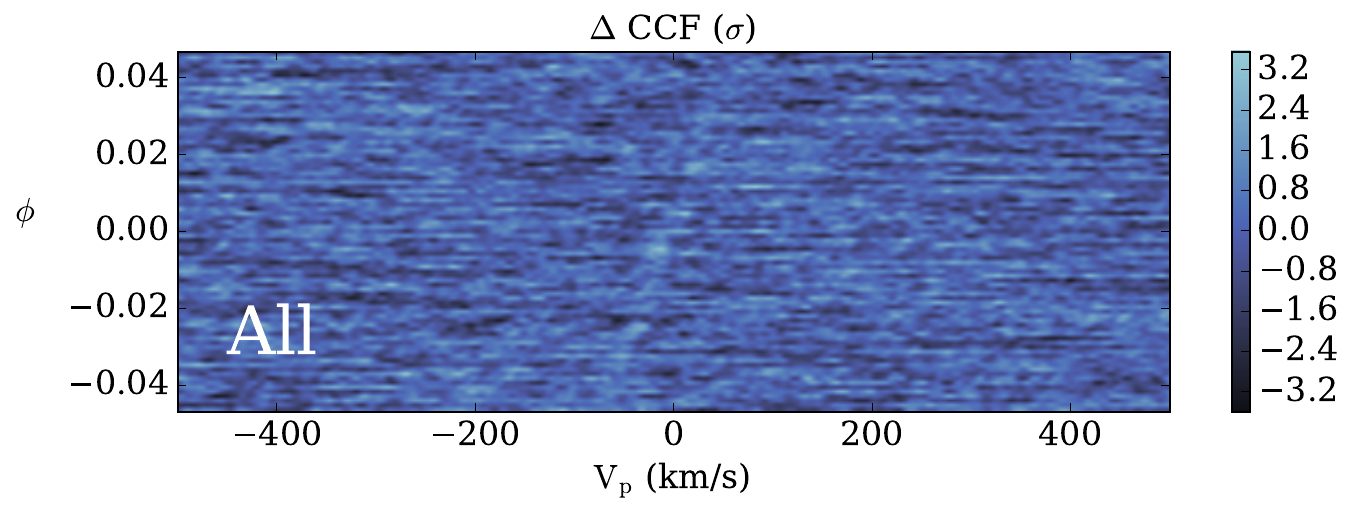}
\end{subfigure}
\caption{Upper: detection significance as a function of $K_p$ and $\Delta V_{sys}$. White dashed lines indicate the literature values and the marker denotes the location of the peak signal. Lower: \correction{Cross correlation value divided by the standard deviation per row, as a function of phase and radial velocity in planet rest frame.} The slope of the trail is approximately $K_p$. Phase-folding the matrix assuming a $K_p$ would yield a single row of the $K_p$-$V_{sys}$ map. Left (reddish) is the cross-correlation result applying emission spectroscopy to WASP-77Ab with a template model containing \ce{H2O, CO}. Right (blueish) is the cross-correlation result applying transmission spectroscopy to WASP-121b with a template model containing all species detected in the retrieval (\ce{Fe, Cr, V, Mg}).}
\label{fig: CCF_both}
\end{figure*}

\begin{figure*}[t]
\centering
\includegraphics[trim=0 0 -50 0, clip, width=0.5\textwidth]{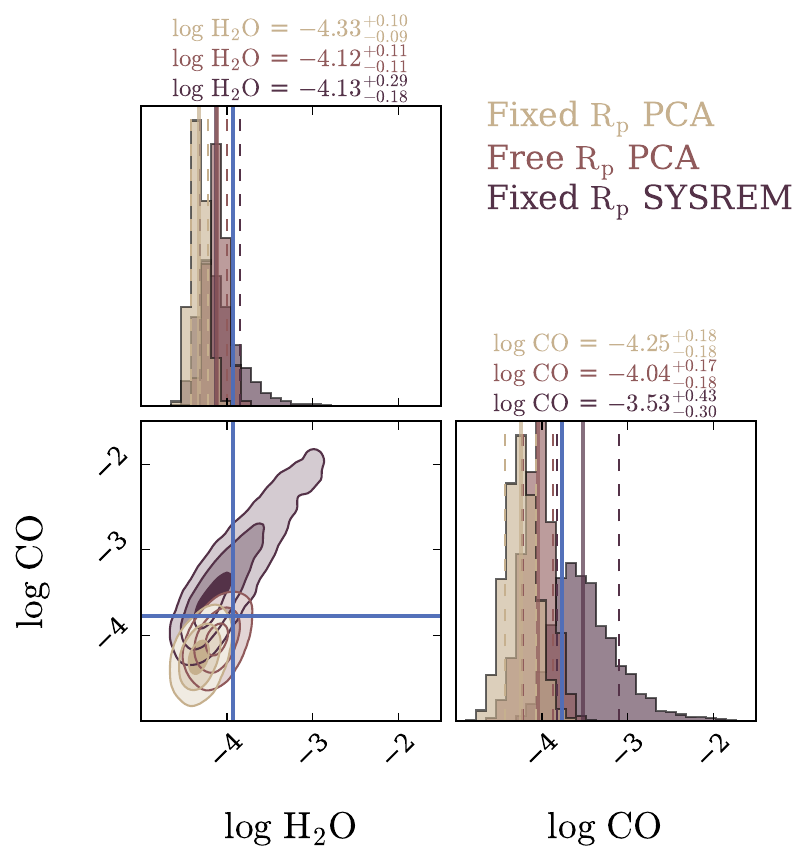}
\includegraphics[width=0.45\textwidth]{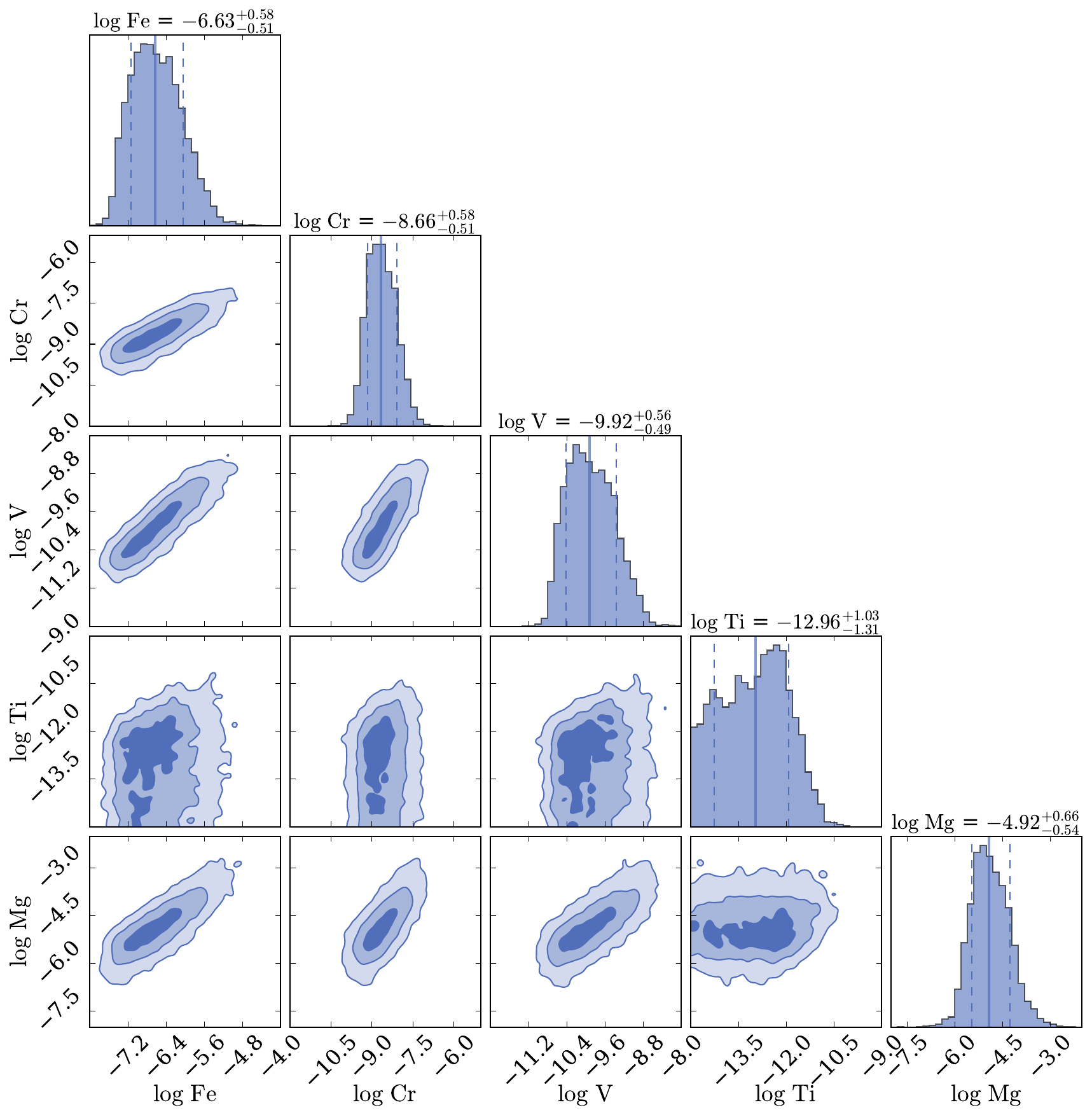}
\caption{Results of the retrieval for the IGRINS WASP-77Ab data. See Figure~\ref{fig: WASP77_combined} for full corner plots. Blue lines are median retrieved values from \cite{Line2021}. Note that fixing planet radius and the choice of filtering methods can slightly offset the results.}
\label{fig: WASP77_corner_combined}
\end{figure*}

Figure~\ref{fig: injection_data_stage} is a visualization of the data at different stages. \correction{To determine the optimal number of iterations for PCA, we test the three different approaches discussed in Section \ref{sec: detrending}. Figure~\ref{fig: CCF_compare} shows the variation of detection significance (S/N) against the number of applied PCA iterations.} We cross correlate template model with \ce{H2O} and \ce{CO} separately, finding strong peak at the expected location in the $K_p\text{-}V_{sys}$ plot (Figure~\ref{fig: WASP77_CCF}), confirming detection of \ce{H2O} and \ce{CO}. We also cross correlate a model with \ce{H2O} and \ce{CO} together and notice that the detection significance becomes stronger as expected (Figure~\ref{fig: CCF_both}). The corner plot with PCA (Figure~\ref{fig: WASP77_corner_combined}) shows the close match between \cite{Line2021} and our results. The slight offsets in retrieved chemical abundances using PCA is possibly due to its degeneracy with planet radius and scale factor $\alpha$ or subtle difference in forward model. Figure~\ref{fig: WASP77_corner_combined} demonstrates that fixing the planet radius and choosing between PCA and SYSREM could affect the results.

Using SYSREM, we also report constrained mixing ratios for $\ce{H2O}$ and \ce{CO} (Figure~\ref{fig: WASP77_combined}). The posteriors are in agreement with \cite{Line2021}. We note the same non-detection of Ti but a spurious detection of $\rm{NH_3}$ in the atmosphere of WASP-77Ab using SYSREM. $\rm{NH_3}$ can be a sign of strong atmospheric mixing as $\rm{NH_3}$ forms deep in the atmosphere in equilibrium. However, the fact that this detection depends on the choice of analysis method makes the detection questionable. We have no concrete evidence of SYSREM outperforming PCA in preserving planet signal. Our proposed explanation is that SYSREM gives less weight to telluric lines than PCA since telluric absorbing regions tend to be more noisy. This makes SYSREM prioritize on fitting the planet signals, more tolerant against alternative models, and provide looser constraints. As high-resolution spectroscopy loses information about the continuum, we view the tight constraint ($\pm 0.1$) on chemical abundances with PCA too optimistic, because we are inferring abundances solely from line depths, whereas other effects that can alter these relative line strengths too such as the temperature structure and clouds.

The detection of \ce{H2O} and \ce{CO} can be seen in Figure~\ref{fig: WASP77_CCF}. We notice the \ce{CO} detection to be weaker than \ce{H2O}. We can barely see the trail in \ce{CO}'s phase-resolved cross correlation function, whereas the trail is much more prominent for \ce{H2O}. These plots are obtained with SYSREM. The stronger detection for \ce{CO} (Figure~\ref{fig: WASP77_CCF}) compared to $\sim 2\sigma$ in \cite{Line2021} hints at SYSREM's improvement in preserving planet signals.

\begin{figure}[t]
\centering
\end{figure}

\section{Application: High-Resolution Transmission Spectroscopy of WASP-121b}
\label{sec: WASP121}

\subsection{Retrieval setting}
We assume isothermal atmosphere and constant-with-altitude volume mixing ratios. We consider the same set of species as \cite{Gibson2022}, which include \ce{Fe}, Cr, V, Ti, Mg. We justify the isothermal assumption through injection tests in Section \ref{sec: injection retrieval setting} and find no motivation for extra P-T parameters as they slow down retrievals. To reproduce the settings of \cite{Gibson2022}, we include a cloud deck with free Rayleigh scattering (controlled by $\gamma$ and $\log a$) \citep{MacDonald2017}. Similar to the injection test (Section \ref{sec: injection retrieval setting}), we use a 100-layer atmosphere equally spaced in log space between $10^{-12}$ bar and 100 bars. We include $\log \alpha$, $K_p$, $V_{sys}$, and $W_{conv}$ as free parameters. We can interpret the retrieved value of this parameter as the degree to which lines are broadened, whether they are astrophysical or instrumental effects. See our discussion in Section \ref{sec: prior} for information about the prior.

\subsection{Results}
A visualization of the data at different stages is shown in Figure~\ref{fig: WASP121_data}. The analysis is done on the combined red and blue data. We cross correlate a number of template models for \ce{Fe, Mg, Cr, V}, and all four species together. Our detection significance shows good match to what's reported in \cite{Merritt2021}. The result in Figure~\ref{fig: CCF_both} is obtained with a model spectrum $\log X_{\ce{Fe}}=-5,\, \log X_{\ce{Mg}}=-4,\, \log X_{\ce{Cr}}=-7,\, \log X_{\ce{V}}=-8$, and $T=3000$ K. Using the best fit model from retrieval in principal and indeed improves detection significance. It is possible to underestimate or miss a detection cross-correlating with a grid of spectra if none of the templates represents the true planet signal well. Therefore, we should be aware that detection significance itself is not a meaningful metric without context. Still, we obtained similar significance using the released code from \cite{Pelletier2023} on this dataset, cross-validating our framework.

We found good match of retrieved temperature, $K_p$, $V_{sys}$, $W_{conv}$, and relative abundance ratios with \citet{Gibson2022}, who reported an isothermal atmosphere of WASP-121b using the same observation. We explored different parametrizations and found that assuming isothermal atmosphere does not affect our results (also see our discussion in Section \ref{sec: forward model}). Retrieved P-T from flexible parametrizations ends up isothermal as well. Our final retrieved temperature T$ = 3587^{+264}_{-238}$ K.

The offset in absolute abundances can be attributed the strong correlation between each chemical abundance and cloud deck pressure. This is expected since transmission spectroscopy is susceptible to strong degeneracy between absolute abundances and cloud properties \citep{Benneke2012, Heng2017}/ \cite{Gibson2022} retrieved atmosphere edges towards the upper prior limit, where \ce{Mg} composes almost all of the gas, appearing to be nonphysical. They do not adjust the mean molecular weight in response to the increased abundances of heavy elements, whereas \POSEIDON do, which could be the source of difference. \cite{Gibson2022} also discuss the possibility of the high abundance of Mg being a signature of atmospheric escape of Mg, since \cite{Sing2019} already provides evidence of \ce{Fe II} and \ce{Mg II} escaping in the atmosphere of WASP-121b. Our retrieved absolute abundances appear more conformed to our knowledge with the $\log X_{\ce{Mg}}$ being well constrained. 

The full corner plot can be found in Figure~\ref{fig: WASP121_corner} in the Appendix. 
 
\section{Summary and Discussion}
\label{sec: summary}

We have presented an open source framework for high-resolution cross-correlation spectroscopy retrievals (HRCCS), for both transmission and emission spectrosa, with the \POSEIDON package. We demonstrated the methods behind HRCCS retrievals and highlighted their potential in atmospheric characterization. We reported small differences in results depending on the filtering algorithm choice. We validated the framework on synthetic data through multiple injection tests. Finally, we applied our high-resolution retrieval framework to IGRINS observation of WASP-77Ab \citep{Line2021} and UVES observation of WASP-121b \citep{Gibson2022}, recovering similar, yet subtly different, atmospheric properties. Notably, we found a spurious detection of NH$_3$ in the atmosphere of WASP-77Ab that arises from the choice of filtering algorithm.

Our new HRCCS retrieval framework offers considerably improved computational speed compared to runtimes mentioned in the literature. A typical emission retrieval costs $\sim 7$ hours on 24 CPUs with 12 parameters and 800 live points, compared to 2.8 days with 24 CPUs and a GPU under the same settings \citep{Line2021}. A typical transmission retrieval costs $\sim 6$ hours on 24 processors for 15 parameters and 800 live points. In comparison, \cite{Klein2024} reports 2-3 hours on 36 processors for 4 parameters and 384 live points, \cite{Pelletier2023} reports 3 months on 18 CPUs with 200 MCMC walkers for 34 parameters. Retrieval runtimes depend on many factors, including the number of MultiNest live points, the number of parameters, the dataset size (number of exposures, wavelength coverage), and the efficiency of CPUs. All computation in this paper is done on an Intel Xeon CPU E5-2670 v3, which came out in 2014. We expect the runtime to be even shorter on a modern CPU (several times faster based on CPU benchmarks online).

Therefore, it is often difficult to compare under the exact same settings. Nevertheless, by using the same datasets and similar model configurations as \citet{Line2021} and \citet{Gibson2022} we see our framework is at least an order of magnitude faster than most HRCCS retrieval runtimes in the literature. Our speed improvements can be attributed to \POSEIDON's optimized forward model, efficient sampling from MultiNest, and our implementation of an efficient fast-filtering algorithm \citep{Gibson2022}.


While we have focused solely on high-resolution retrievals, it is straightforward to generalize to combined high and low-resolution retrievals. For example, \cite{Gandhi2019} and \citet{Smith2023} have demonstrated retrievals combining the strengths of both HRS and LRS observations to probe complementary aspects of hot Jupiter atmospheres. Our fast, easy-to-use, open source, high-resolution retrieval framework for emission and transmission spectra can be easily coupled with low-resolution data and thereby empower the exoplanet community to realize new atmospheric perspectives for a myriad of exoplanets. 

Future work can address constraints on multidimensional atmospheric properties from high-resolution ground-based spectra. Recent studies have demonstrated HRS can reveal multi-dimensional structures with phase-resolved cross-correlation function \citep[e.g.][]{Ehrenreich2020, Gandhi2022, Pelletier2023}. Since our retrieval framework inherits the multidimensional transmission spectra modeling capabilities of \POSEIDON, our code is already equipped with multidimensional HRCCS retrieval functionality. Given the fast runtime, we find for 1D models, future HRCCS retrieval studies with \POSEIDON can readily investigate the retrievability of multidimensional atmospheric properties. 

High-resolution spectroscopy has proven prolific in detecting chemical species in hot Jupiters and has recently begun to offer statistical constraints through HRCCS retrievals. With a plethora of ground-based high-resolution instruments providing an abundance of high-quality datasets and a new generation of telescopes on the horizon, we hope that our open source HRCCS retrieval framework will catalyze new insights into hot Jupiter atmospheres.


\newpage

\section*{Acknowledgments}

R.J.M. is supported by NASA through the NASA Hubble Fellowship grant HST-HF2-51513.001, awarded by the Space Telescope Science Institute, which is operated by the Association of Universities for Research in Astronomy, Inc., for NASA, under contract NAS 5-26555. We extend gratitude to Michael Line for publicly providing an example code on Dropbox for high-resolution retrievals. We thank Jake Turner for sharing his encyclopedic knowledge of high-resolution observing facilities.

\section{Data availability}

All data products and scripts to reproduce these results will be available upon request. \correction{The high-resolution retrieval framework is part of \POSEIDON Release 1.3.}

\bibliographystyle{aasjournal}
\bibliography{export-bibtex}

\begin{appendix}

\begin{table}[h]
\centering
\begin{tabular}{ccc}
\hline \noalign{\smallskip}
Parameter [units] & Injected value & Prior \\\noalign{\smallskip}
\hline \noalign{\smallskip}
\multicolumn{3}{l}{\it Emission Injection Test:} \\\noalign{\smallskip}
~$\log \alpha$             & $0$ & $\mathcal{U}(-2, 2)$ \\ \noalign{\smallskip}
~$K_{\rm p}$ [km/s]          & $-200$ & $\mathcal{U}(-300,-100)$  \\ \noalign{\smallskip}
~$v_{\rm sys}$ [km/s] & $-20$ & $\mathcal{U}(-50,50)$  \\ \noalign{\smallskip}
~$W_{\rm conv}$      & $1$ & $\mathcal{U}(0.1, 20)$ \\\noalign{\smallskip}
~$\log_{10}(X_{\rm species})$ & $-4$ & $\mathcal{U}(-15, 0)$ \\\noalign{\smallskip}
~$T_{\rm P=10^{-2} bar}$  [K]      & 1700 & $\mathcal{U}(400, 4000)$  \\\noalign{\smallskip}
~$a_1$              & 0.2 & $\mathcal{U}(0.02,1)$  \\\noalign{\smallskip}
~$a_2$               & 0.1 & $\mathcal{U}(0.02,1)$ \\\noalign{\smallskip}
~$\log P_1$           & 0.17 & $\mathcal{U}(-5,2)$ \\\noalign{\smallskip}
~$\log P_2$              & -1.39 & $\mathcal{U}(-5,2)$ \\\noalign{\smallskip}
~$\log P_3$              & 1 & $\mathcal{U}(-2,2)$  \\\noalign{\smallskip}
\hline \noalign{\smallskip}
\multicolumn{3}{l}{\it Transmission Injection Test 1:} \\\noalign{\smallskip}
~$R_{\rm r, ref}$ [$R_J$]       & $  1.75 $ & $\mathcal{G}(1.75,0.05)$ \\\noalign{\smallskip}
~$\log \alpha$                  & $ 0 $ & $\mathcal{U}(-2,2)$ \\\noalign{\smallskip}
~$K_{\rm p}$ [km/s]             & $ -200$ & $\mathcal{U}(-300,-100)$ \\\noalign{\smallskip}
~$v_{\rm sys}$ [km/s]           & $    -20$ & $\mathcal{U}(-50,50)$ \\\noalign{\smallskip}
~$W_{\rm conv}$      & $1$ & $\mathcal{U}(0.1, 20)$ \\\noalign{\smallskip}
~$\log_{10}(X_{\rm{species}})$    & $   -6$ & $\mathcal{U}(-15,0)$ \\\noalign{\smallskip}
~$T_{\rm P=10^{-2} bar}$  [K]      & 3000 & $\mathcal{U}(400,4000)$  \\\noalign{\smallskip}
~$a_1$              & 0.1 & $\mathcal{U}(0.02,2)$   \\\noalign{\smallskip}
~$a_2$               & 0.2 & $\mathcal{U}(0.02,2)$  \\\noalign{\smallskip}
~$\log P_1$           & -1 & $\mathcal{U}(-5,2)$   \\\noalign{\smallskip}
~$\log P_2$              & -2 & $\mathcal{U}(-5,2)$   \\\noalign{\smallskip}
~$\log P_3$              & 2 & $\mathcal{U}(-2,2)$   \\\noalign{\smallskip}
\hline \noalign{\smallskip}
\multicolumn{3}{l}{\it Transmission Injection Test 2:} \\\noalign{\smallskip}
~$T$ [K]  & - & $\mathcal{U}(2000,4000)$ \\\noalign{\smallskip}
\hline
\noalign{\smallskip}
\end{tabular}
\caption{Parameters for the two injection tests. $\mathcal{U}(a,b)$ represents a uniform (improper) prior, where the prior is zero if the parameter is outside the limits, and equal to one otherwise. See text for explanations.}
\label{tab:injectiontests}
\end{table}

\begin{figure}[ht]
\centering
\begin{subfigure}
  \centering
  \includegraphics[trim=20 10 10 0, clip, width=0.85\linewidth]{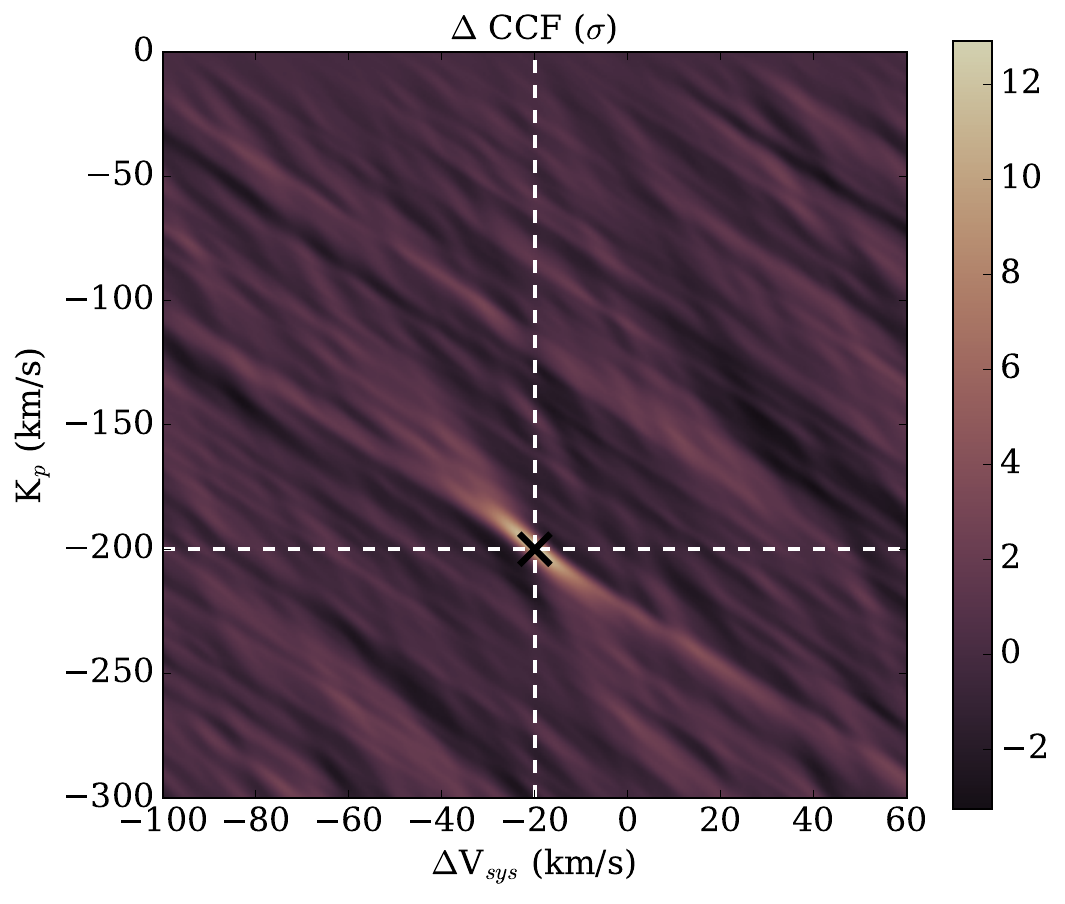}
\end{subfigure}
\begin{subfigure}
  \centering
  \includegraphics[width=0.8 \linewidth]{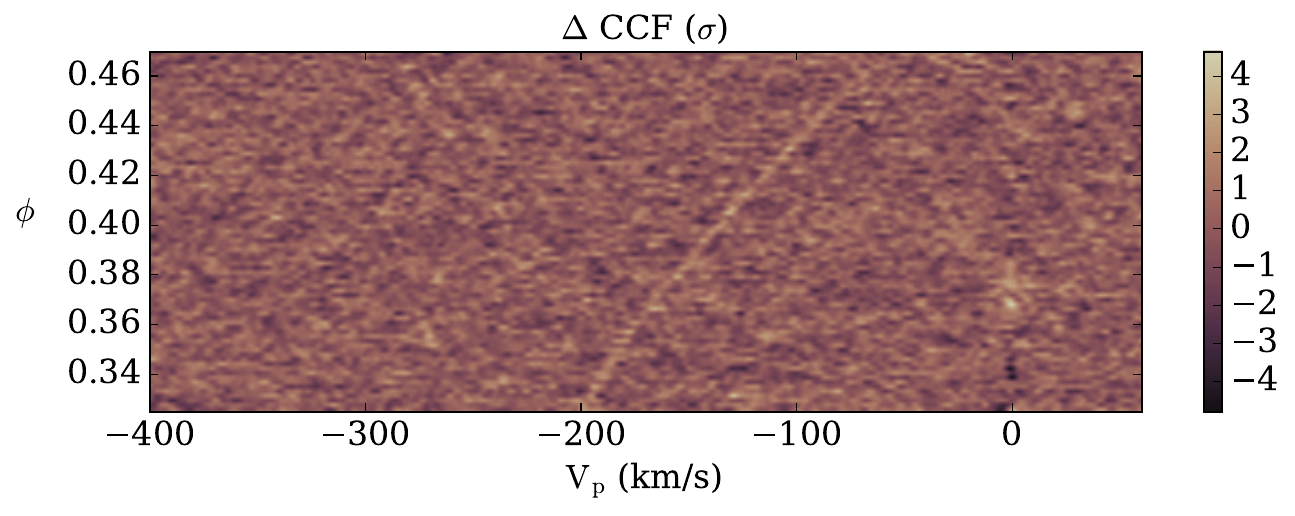}
\end{subfigure}
\begin{subfigure}
  \centering
  \includegraphics[trim=20 0 10 0, clip, width=0.85 \linewidth]{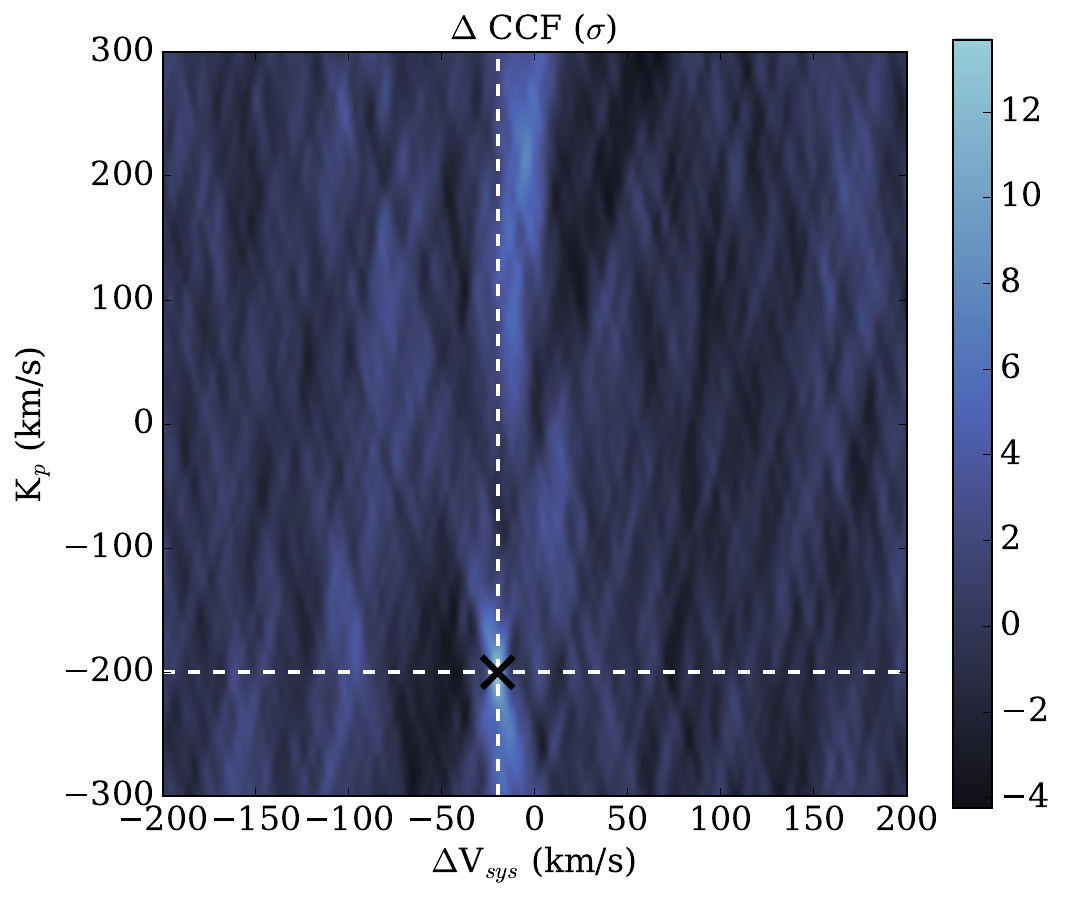}
\end{subfigure}
\begin{subfigure}
  \centering
  \includegraphics[width=0.85 \linewidth]{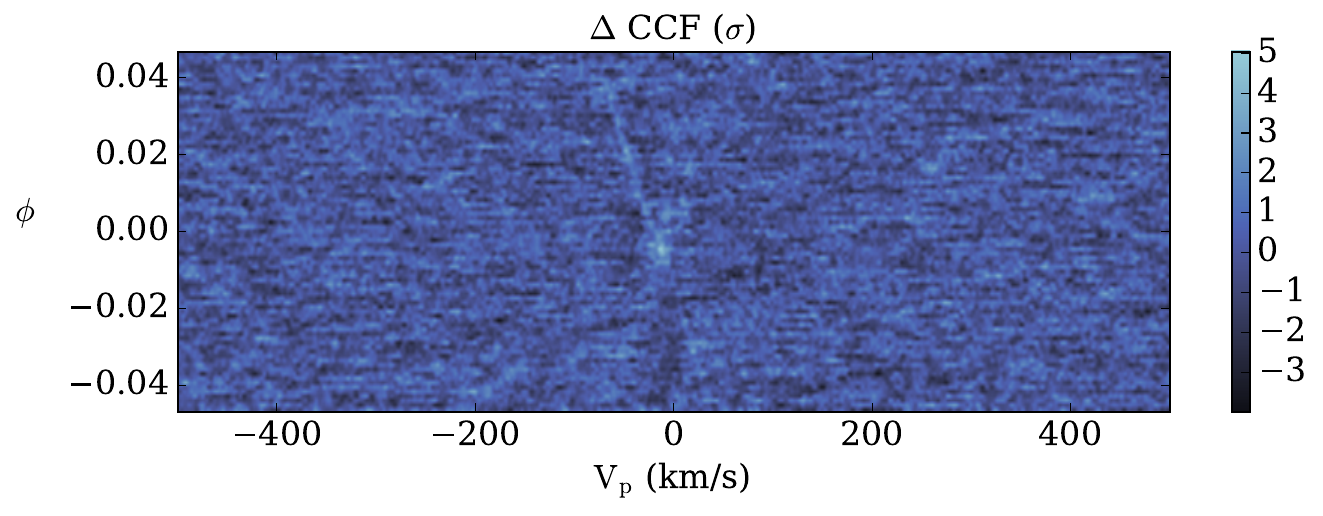}
\end{subfigure}
\caption{Detection significance of an injected signal as a function of $K_p$ and $\Delta V_{sys}$ and detection significance as a function of phase and planet radial velocity (normalized by standard deviation per phase). Red: emission. Blue: transmission.}
\label{fig: injection_CCF}
\end{figure}

\begin{figure*}[ht]
\centering
\includegraphics[width=\textwidth]{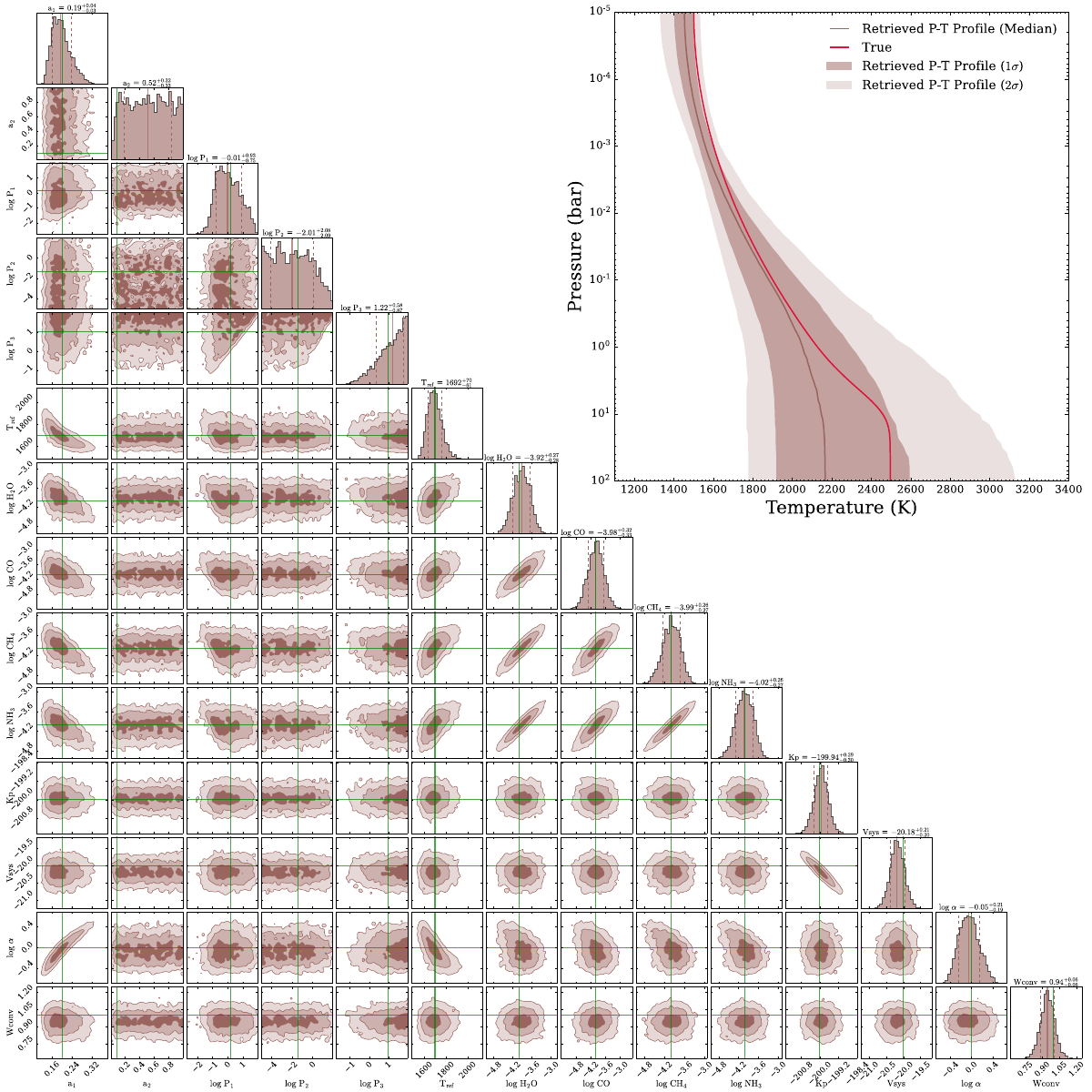}
\caption{Full corner plot of retrieval results on injected data ($\log X_{\ce{H2O}}=\log X_{\ce{CO}}=\log X_{\ce{CH4}}=\log X_{\ce{NH3}}=-4$) along with retrieved P-T profile using SYSREM. Run time: 13 hours on 24 cores.}
\label{fig: WASP77_injection_H2O_CO_CH4_NH3_combined}
\end{figure*}

\begin{figure*}[ht]
\centering
\includegraphics[width=\textwidth]{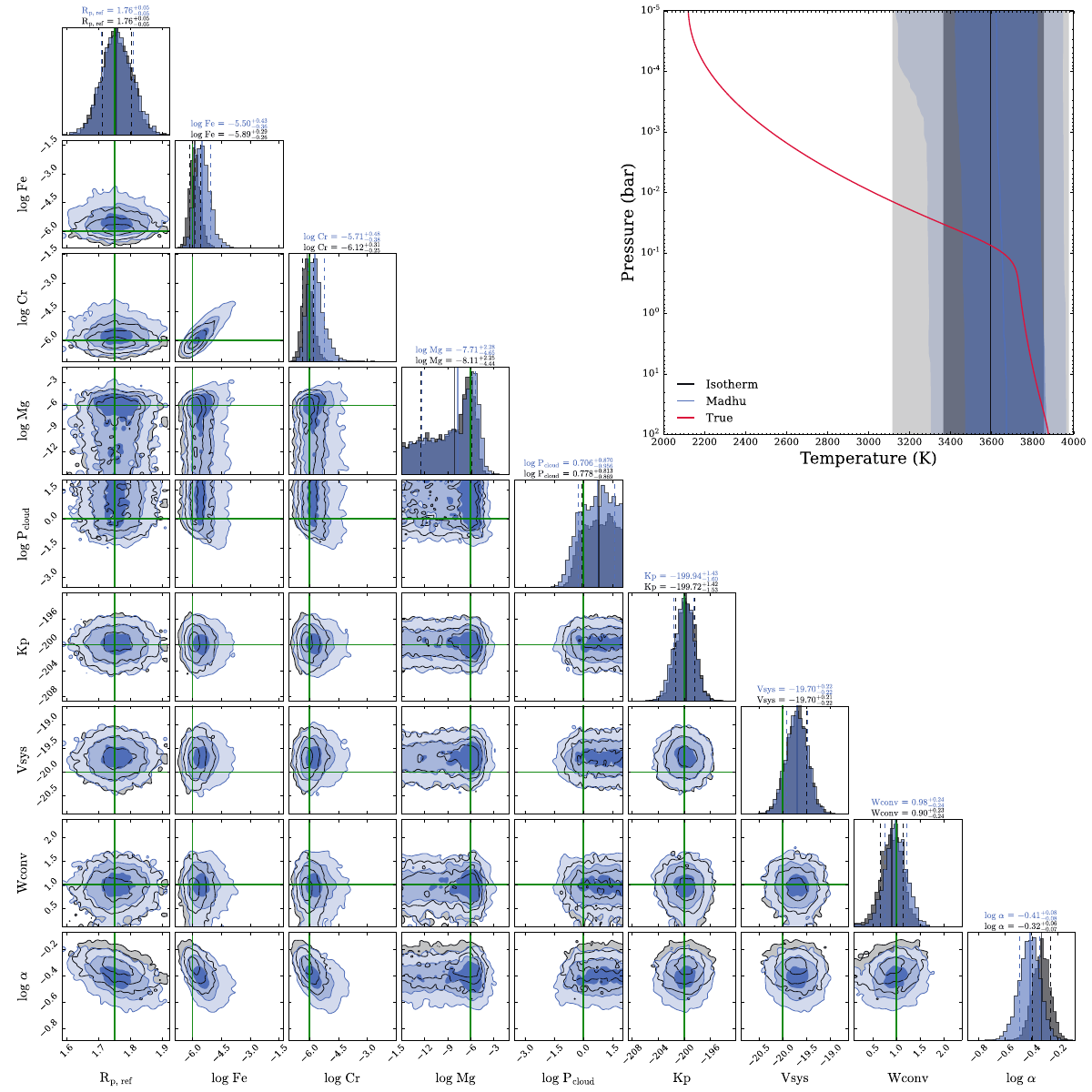}
\caption{Full corner plots of retrieval results on simulated WASP-121b data. The PT profile of the injection signal is parameterized by six parameters. Blue: retrieval assuming isothermal atmosphere. Red: retrieval with six-parameter P-T profile. Upper right: true P-T profile compared to retrieved P-T profile. Theses results show high-resolution retrievals are not particularly sensitive to temperature gradients. Run time: both about 6 hours on 24 cores.}
\label{fig: WASP121_injection_all_retrieved}
\end{figure*}

\begin{figure*}[ht]
\centering
\includegraphics[width=\textwidth]{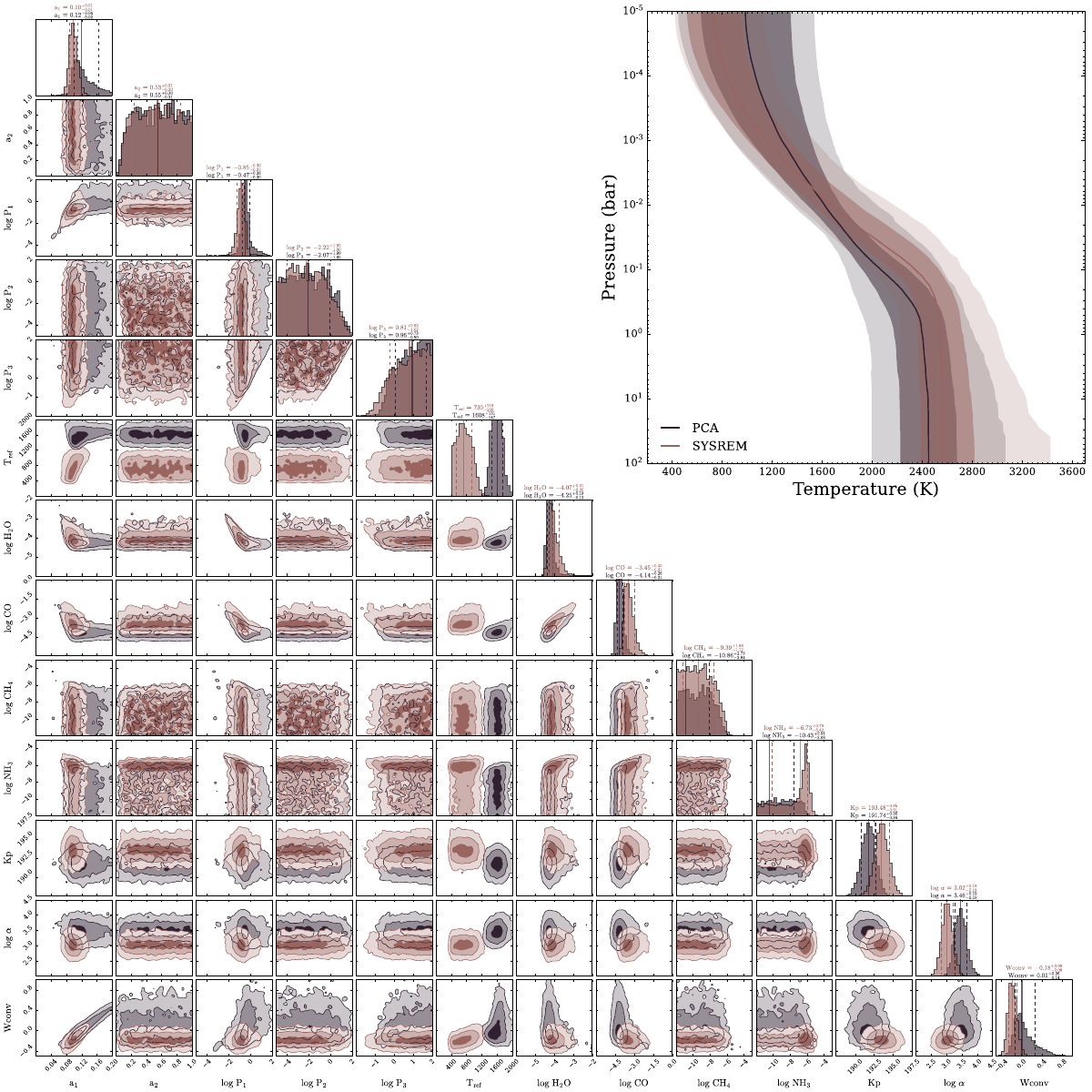}
\centering
\caption{Full corner plot of the retrievals on IGRINS WASP-77Ab data along with retrieved P-T profile. Blue: PCA; run time: 11 hours on 24 cores. Red: SYSREM; run time: 6.6 hours on 24 cores.}
\label{fig: WASP77_combined}
\end{figure*}

\begin{figure*}[ht]
\centering
\includegraphics[width=\textwidth]{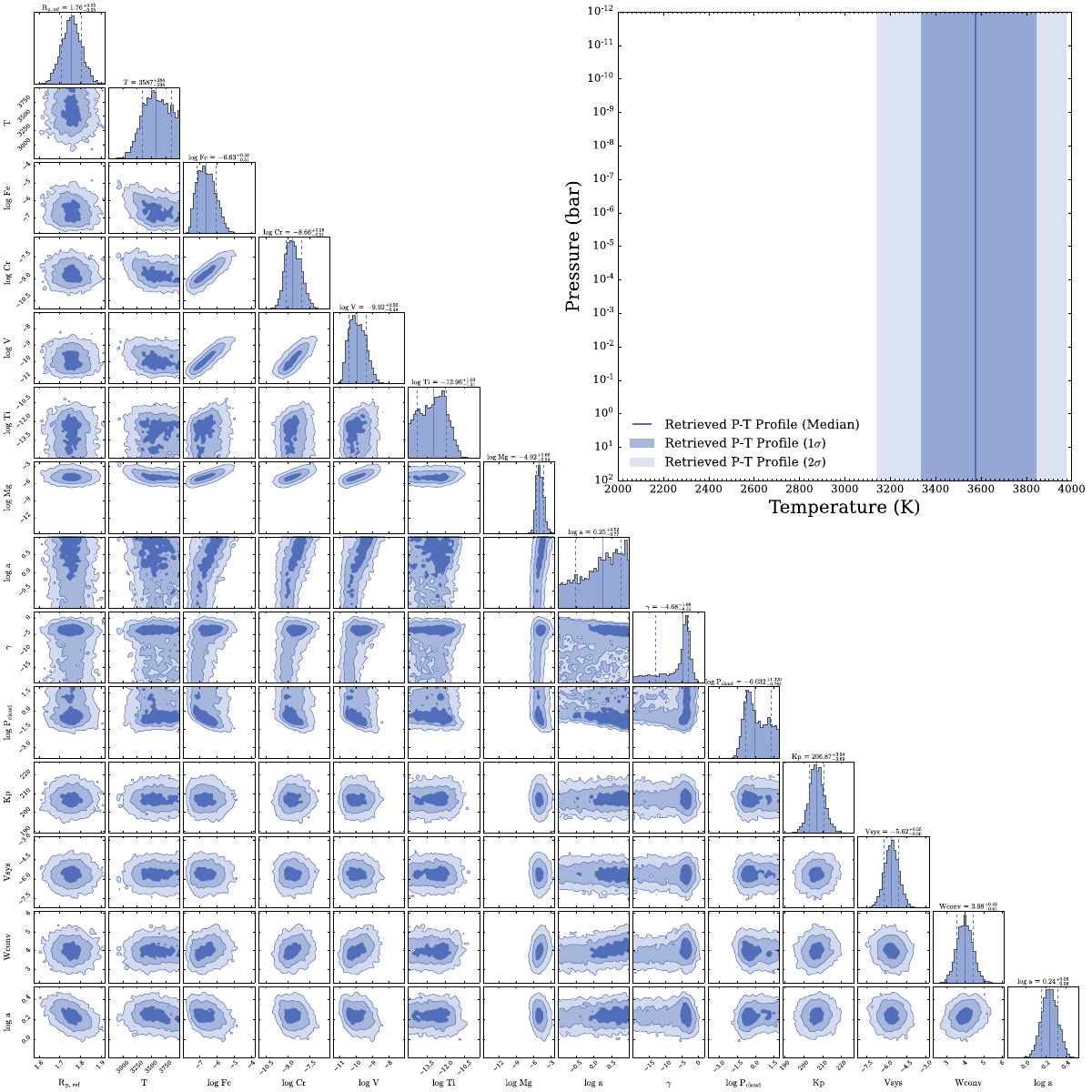}
\caption{Full corner plot of the retrieval for the combined red and blue WASP-121b data along with retrieved P-T profile. Run time: 7.3 hours on 24 cores.}
\label{fig: WASP121_corner}
\end{figure*}

\begin{figure*}[t]
\vspace{-7.5cm}
\centering
\begin{subfigure}
  \centering
  \includegraphics[width=0.45\linewidth]{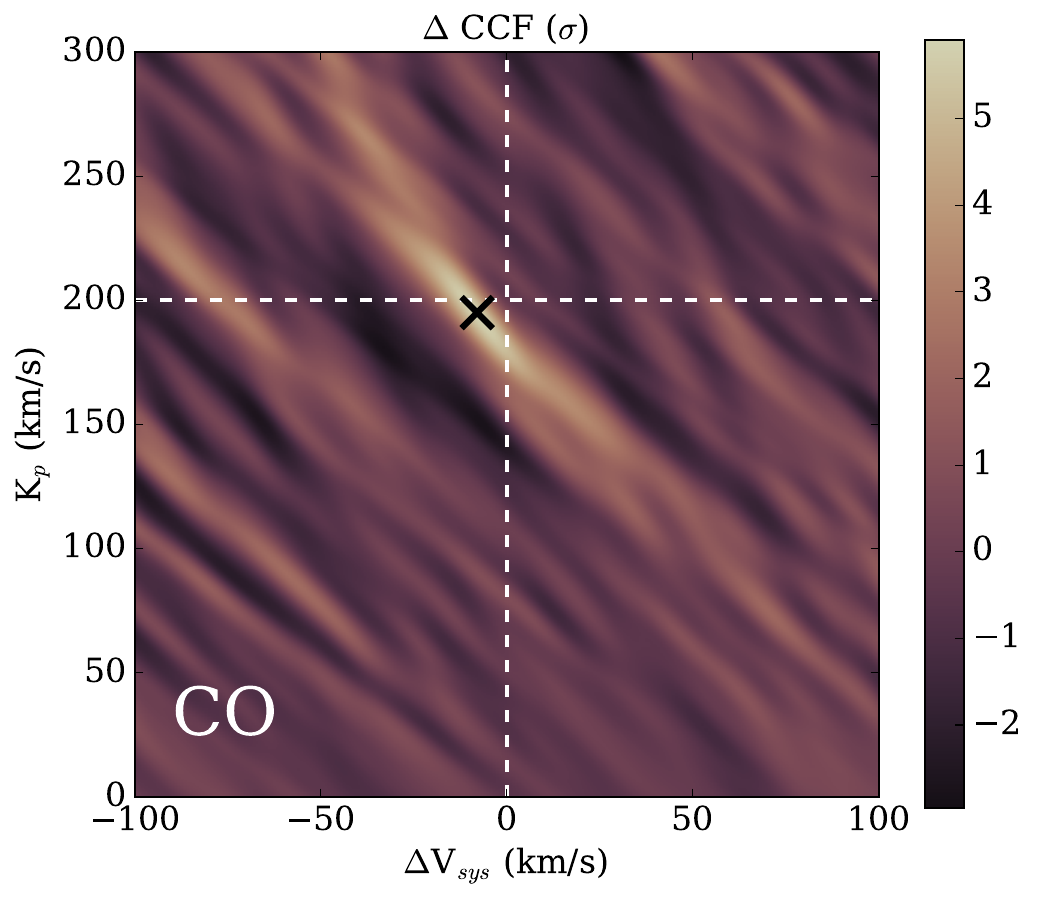}
\end{subfigure}
\begin{subfigure}
  \centering
  \includegraphics[width=0.45\linewidth]{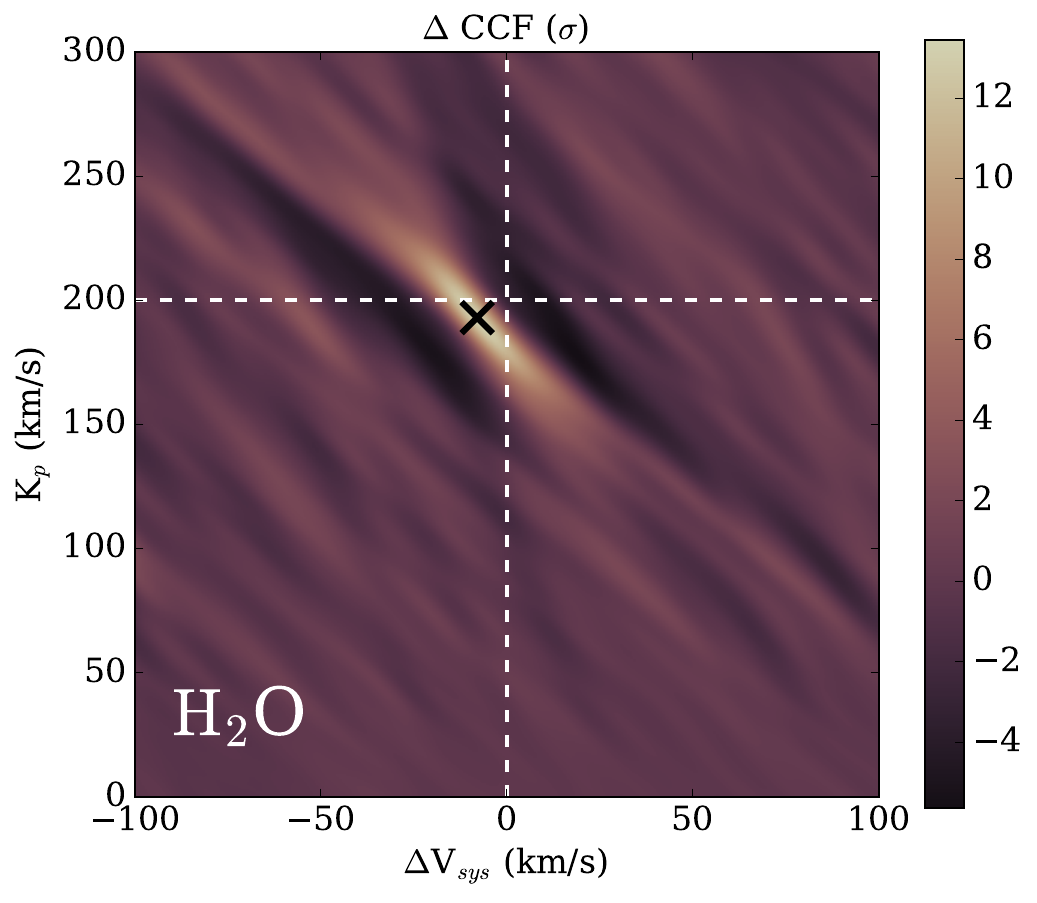}
\end{subfigure}
\begin{subfigure}
  \centering
  \includegraphics[width=0.45\linewidth]{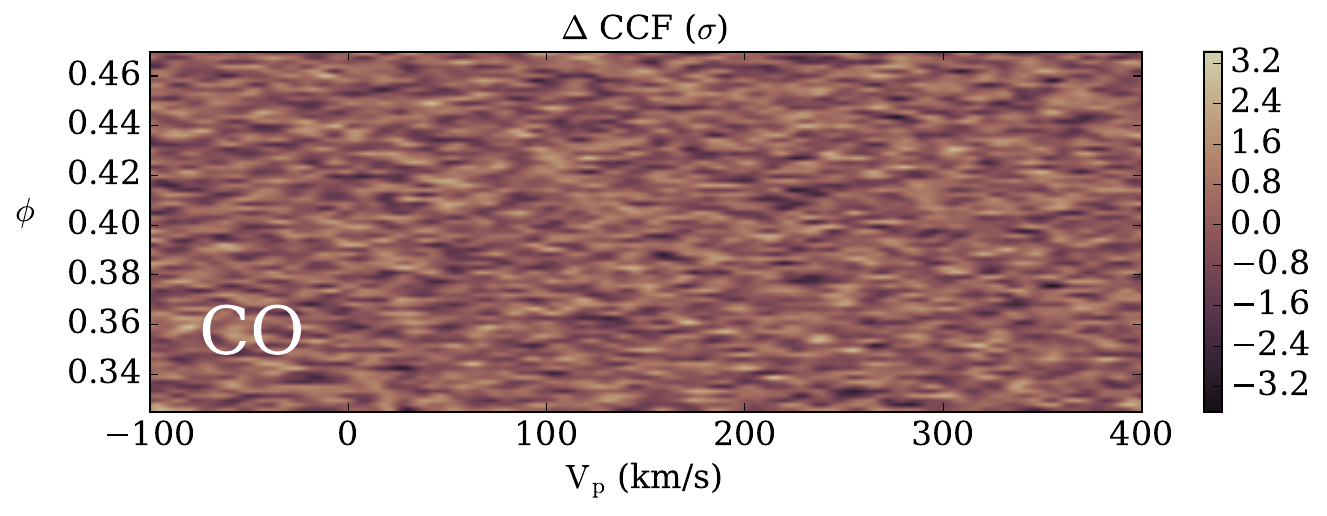}
\end{subfigure}
\begin{subfigure}
  \centering
  \includegraphics[width=0.45\linewidth]{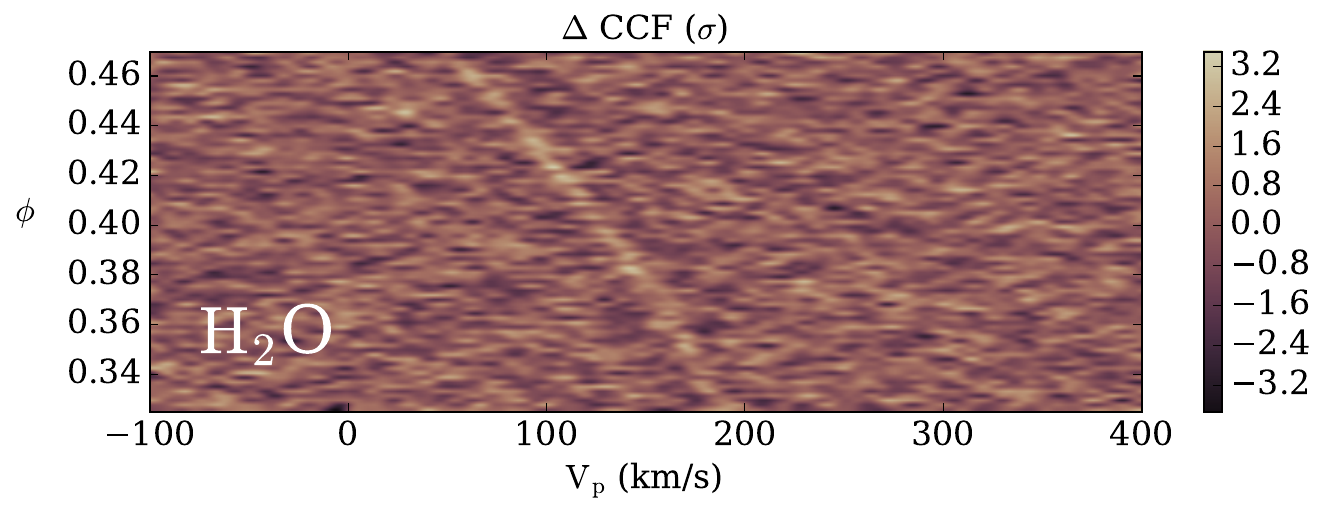}
\end{subfigure}
\caption{Cross-correlation results with \ce{CO} and \ce{H2O} separately for WASP-77Ab. White dashed lines are approximately the published values.}
\label{fig: WASP77_CCF}
\end{figure*}

\end{appendix}
\end{document}